\documentclass[prd,nofootinbib]{revtex4}
\usepackage[latin1]{inputenc}
\usepackage{pstricks,pst-node,pst-text,pst-3d,pslatex}
\usepackage[T1]{fontenc}

\usepackage{graphicx}
\usepackage{amsfonts}
\usepackage{amsmath}

\newcommand{\ox}{\bar{x}}
\newcommand{\be}{\begin{eqnarray}}
\newcommand{\ee}{\end{eqnarray}}

 \newcommand{\om}{\omega}
\newcommand{\Om}{\Omega}
\newcommand{\mmp}{\varDelta t_{m+1,m}}
\newcommand{\mmn}{\varDelta t_{m,m-1}}
\newcommand{\veff}{V_{eff}}    

\begin{document}

\title{ Fluctuations in the Ensemble of Reaction Pathways} 
\author{G. Mazzola}
\affiliation{Dipartimento di Fisica  Universit\'a degli Studi di Trento, Via Sommarive 14, Povo (Trento), I-38050 Italy. 
International School for Advanced Studies (SISSA), via Bonomea 265
34136 Trieste, Italy}
\author{S. a Beccara}
\affiliation{Dipartimento di Fisica  Universit\'a degli Studi di Trento, Via Sommarive 14, Povo (Trento), I-38050 Italy.}
\affiliation{INFN, Gruppo Collegato di Trento, Via Sommarive 14, Povo (Trento), I-38050 Italy.} 
\author{P. Faccioli}
\affiliation{Dipartimento di Fisica  Universit\'a degli Studi di Trento, Via Sommarive 14, Povo (Trento), I-38050 Italy.}
\affiliation{INFN, Gruppo Collegato di Trento, Via Sommarive 14, Povo (Trento), I-38050 Italy.} 
\email{faccioli@science.unitn.it}
\author{H. Orland}
\affiliation{Institut de Physique Th\'eorique,
Centre d'Etudes de Saclay, F-91191, Gif-sur-Yvette, France}

\begin{abstract}

The dominant reaction pathway (DRP) is a rigorous framework to microscopically compute the most
probable trajectories, in  non-equilibrium transitions. 
In the low-temperature regime, such dominant pathways encode the information about the reaction mechanism and can be used to estimate non-equilibrium averages of arbitrary 
observables.
On the other hand, at sufficiently high temperatures, the stochastic fluctuations around the dominant paths become important and have to be taken into account. 
In this work, we develop a technique to systematically include the effects of such stochastic fluctuations, to order $k_B T$. This method is used to compute the probability for a transition to take place through
a specific reaction channel and to evaluate the reaction rate. 
\end{abstract}

\maketitle

\section{Introduction}

The theoretical investigation of the kinetics of rare conformational reactions of macromolecules represents a fundamental, yet very challenging task. In view of the large computational cost of  performing 
Molecular Dynamics (MD) simulations of the long-time dynamics of molecular systems, alternative methods have been developed, which allow to sample directly the space of the reactive trajectories in large conformational spaces, without investing time in simulating thermal oscillations in the (meta)-stable states\cite{TPS1, TPS2, Elber, Donniach, DIMS, elber1, DRP1, DRP2, DRP3}. 

In particular, the Dominant Reaction Pathways (DRP)  approach~\cite{elber1, DRP1,DRP2, DRP3}  yields by construction the set of \emph{most statistically significant} reactive trajectories in the over-damped limit of Langevin dynamics. 
In such an approach, the reactive paths are not calculated by integrating the equation of motion of the system.  Instead, they are obtained by minimizing a target functional, which is rigorously derived starting from the original Langevin equation.

The main advantage of the DRP approach is that it allows to remove the time as the independent variable. Instead,  the dominant path is calculated as a function of a curvilinear abscissa $l$ which measures the distance covered in configuration space. This way,  the problem of the decoupling of the time scales in the internal dynamics of molecular systems is rigorously bypassed and typically  $o(100)$ path discretization steps are 
sufficient to characterize an entire conformational transition. 
In the same formalism, the time-dependent dominant pathway $x(t)$  can be rigorously calculated \emph{a posteriori}, from the trajectory $x(l)$.

\begin{figure}[t]
\includegraphics[width=12 cm]{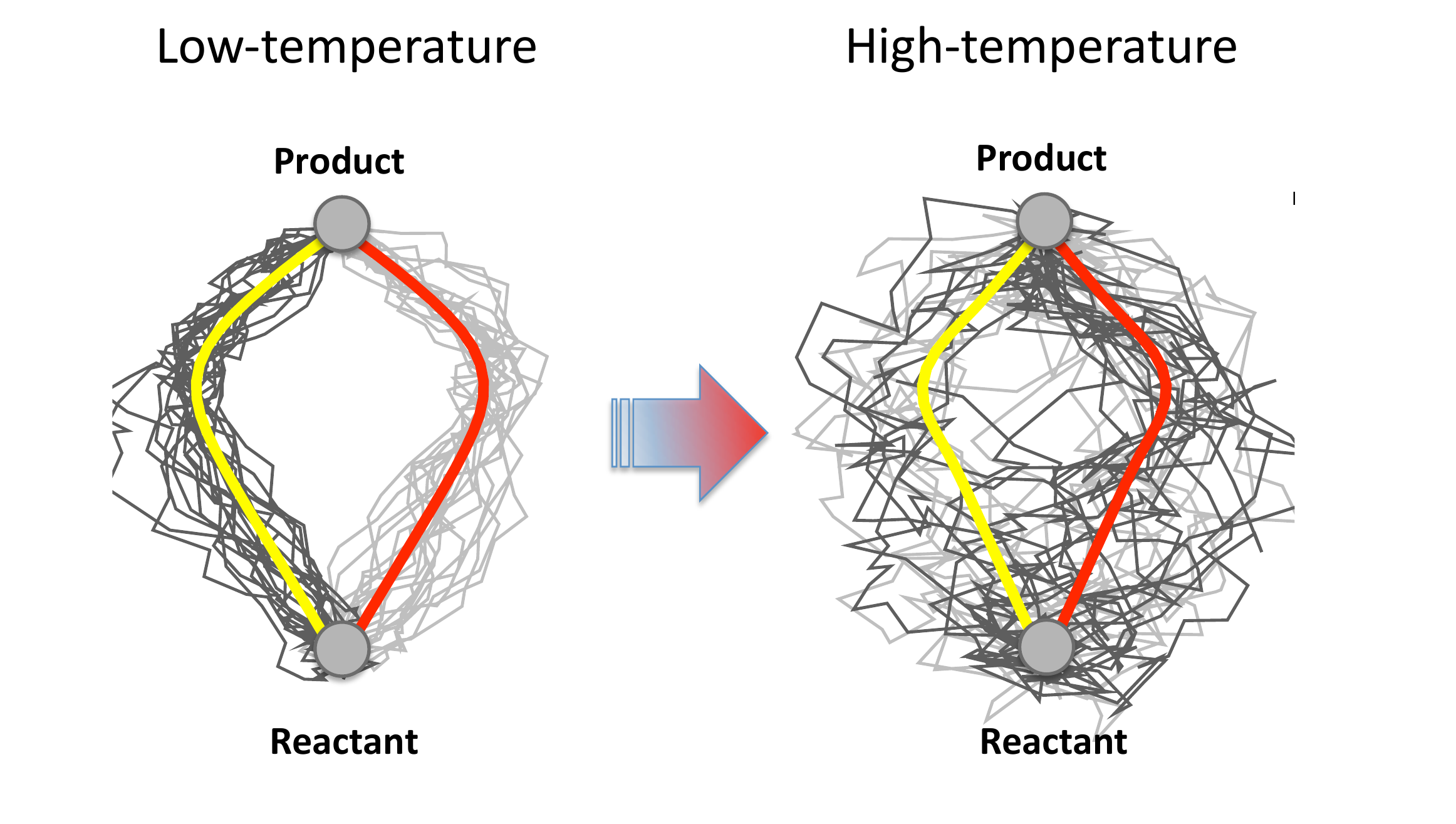}
\caption{Stochastic fluctuations around two different dominant reaction pathways in a two-dimensional transition. As the temperature raises, the stochastic fluctuations become large and eventually
the two paths become statistically indistinguishable.   }
\label{fig1}
\end{figure}

A potential limitation of the DRP approach is that it emphasizes the role played the \emph{most probable} reactive  trajectories. These are smooth paths, defined as the maxima of the functional probability density $\mathcal{P}[x(l)]$ for the system to make a transition  from given  reactant  to given product configurations.
Clearly, the real physical transitions never occur through such smooth dominant paths, but only through non-differentiable stochastic trajectories.  However, it is possible to show that in 
the low-temperature 
limit the path probability density $\mathcal{P}[x]$ is   peaked in the regions of the functional path space surrounding the  dominant reaction pathways. 
In such a temperature regime, each dominant reaction pathway can be considered as representative of a different reaction channel ---see Fig. \ref{fig1}---.  Hence, if the reaction can occur through $n$ different channels ---i.e. there are $n$ distinct dominant reaction pathways--- then the probability of making a transition through the $i-$th channel can be estimated from  the equation 
\be
\label{LO}
Prob.(\textrm{i-th reaction channel }) \simeq  \frac{\mathcal{P}[\bar x_i(l)]}{\sum_{k=1}^n~\mathcal{P}[\bar x_k(l)]},
\ee
where $\mathcal{P}[\bar x_k(l)]$ is the probability density of the $k-$th dominant reaction pathway, $\bar x_k(l)$.   
In this formula, the stochastic fluctuations  around the dominant paths are completely neglected. 

The amplitude of the stochastic fluctuations grows with the temperature of the heat-bath --- see Fig. \ref{fig1}---.
In general, we expect that in  many practical applications it is necessary to go beyond the  leading-order approximation (\ref{LO}), i.e. to  account also for the effect of  small stochastic fluctuations around the dominant paths through an expansion in the thermal energy $k_B T$. 
As we shall see, to  the next order in $k_B T$ the probability of a reaction channel  can be cast in the form
\be
\label{NLO}
Prob.(\textrm{i-th reaction channel }) \simeq  \frac{f[\bar x_i] \mathcal{P}[\bar x_i]}{\sum_{k=1}^n~f[\bar x_k]~\mathcal{P}[\bar x_k]},
\ee
where $f[\bar x_k]$ is the $o(k_B T)$ correction to the probability of the reaction channel defined by the $k$-th dominant path.   
 
The purpose of the present paper is to study the role of such stochastic fluctuations in the reaction kinetics. 
In the first part of the work, we provide a rigorous and practical method to compute the $f[\bar x_k]$ coefficients, i.e. the contribution to the \emph{normalized} path probability $\mathcal{P}[x]$
arising from the path integral over the small stochastic fluctuations around the most probable paths, to order $k_B T$.  In general,  numerically performing such a calculation can be  quite  computationally demanding. 
However, we shall see that using a recently developed formulation of the Langevin dynamics at  low-time resolution power  \cite{EST, RGMD} it is possible to drastically reduce its computational cost. 

In the second part of this work,  we derive an expression which relates the reaction rate to the probability densities in configuration space,  evaluated  in the DRP approach to order $k_B T$. In order, to illustrate 
and test our method 
with a high numerical accuracy, in this work we choose to focus on thermally activated transitions in simple low-dimensional toy systems and leave the analysis of 
much more complicated molecular reactions to our future work.  We shall  show that, once thermal fluctuations around the dominant paths are correctly taken into account, it is possible to obtain accurate estimates for the reaction rates and equilibrium population ratios.

The paper is organized as follows. In section \ref{PIreview} and \ref{DRPform} we review the DRP formalism. 
In the subsequent section \ref{pra} we show how to compute the contribution to the reaction path probability arising form small stochastic fluctuations around the dominant paths. 
In section \ref{convergence}, we show that the 
efficiency of such a calculation can be greatly improved by adopting the low-time resolution formulation of the Langevin dynamics, which was recently developed in \cite{EST, RGMD}.  
In section \ref{test1}, we test our calculation of the path probability in an analytically solvable model and we find that it gives very accurate results.  In section \ref{rates} we derive an expression for
the reaction rates in terms of the path probability density calculated in the DRP approach, while in section \ref{testrate} we test this formula by computing the reaction rate in some  test systems. Our results and conclusions are summarized in section \ref{conclusions}.

 \section{Path Integral Representation of  the Over-damped Langevin Dynamics}
 \label{PIreview}
 

Let us  consider a system defined on a generic $d$-dimensional configuration space. We shall assume that the dynamics  is defined by the over-damped Langevin equation
\be
\label{Langevin}
\dot x = - \beta D \nabla U(x) + \eta(t),
\ee  
where $x$ is a point in the configuration space, $\beta=1/k_B T$, $D$ is the diffusion coefficient, $U(x)$ is the potential energy function and $\eta(t)$ is delta-correlated Gaussian noise, satisfying the fluctuation-dissipation relationship:
\be
\langle \eta^i(t') \eta^j(t) \rangle = 2 \delta^{i j} D \delta(t-t')\qquad (i, j = 1, d).
\ee
Note that in the original Langevin Eq. there is a mass term $m \ddot x$. However, for macro-molecular systems this term is damped at a time scale $10^{-13}~s$, which much smaller than the time scale associated to local conformational changes.

The stochastic differential Eq. (\ref{Langevin}) generates a probability distribution $P(x,t)$ which obeys the well-known Smoluchowski Eq.:
\be
\label{FP}
\frac{\partial}{\partial t} P(x,t) = D \nabla \left[ \nabla P(x,t) + \beta \nabla U(x) P (x,t)\right].
\ee
Such an Eq. is often written in the form of a continuity equation,
\be
\frac{\partial}{\partial t} P(x,t)= - \nabla J(x, t),
\ee
where 
\be
\label{Jdef}
J(x,t) = -D \left[ \nabla  + \beta \nabla U(x) \right]~P(x,t)
\ee
is the so-called probability current.

By performing the formal substitution 
$
P(x,t)= e^{-\frac{\beta}{2 } U(x)}~\Psi(x,t),
$
the Smoluchowski Eq. (\ref{FP}) can be recast in the form of an imaginary time  Schr\"odinger Eq.:
\be
 -\frac{\partial}{\partial t} \Psi(x,t) = \hat{H}_{eff}~\Psi(x,t),
\label{SE}
\ee
where
\be
\hat{H}_{eff}~=~- D \nabla^2 + V_{eff}(x),
\label{Heff}
\ee
  is an effective "quantum" Hamiltonian operator and  
\be
V_{eff}(x)= \frac{\beta^2 D}{4}\left( (\nabla U(x))^2- \frac 2 \beta \nabla^2 U(x)\right).
\label{Veff}
\ee
is called the effective potential.

The conditional probability $P(x_f, t| x_i)$  to find the system at the configuration $x_f$ at time $t$, provided it was prepared in the configuration $x_i$ at time $t=0$ is the Green's function of the 
Smoluchowski Eq., i.e. 
\be
\frac{\partial}{\partial t}  P(x_f,t|x_i,0)-D \nabla \left[ \nabla  P(x_f,t|x_i,0) + \beta \nabla U(x_f)  P(x_f,t|x_i,0)\right]= \delta(x_f-x_i) \delta(t).
\ee

Formally, the  conditional probability $P(x_f, t| x_i)$ can be related to the imaginary time   propagator of the effective "quantum" Hamiltonian (\ref{Heff}):
\be
P(x_f,t|x_i) &=&  e^{-\frac{\beta}{2} (U(x_f)-U(x_i))} ~K(x_f,t|x_i) =~e^{-\frac \beta 2 (U(x_f)-U(x_i))} ~   \langle x_f | e^{- t H_{eff} }| x_i\rangle.
\label{K}
\ee
Using such a connection, it is immediate to obtain an expression of  the conditional probability (\ref{K}) in the form of a Feynman path integral
\be
\label{PI}
P(x_f,t|x_i) = e^{-\frac \beta 2 (U(x_f)-U(x_i))} ~\mathcal{N}~\int_{x(t_i)=x_i}^{x(t)=x_f} \mathcal{D} x~e^{-\int_{0}^t d\tau ~\left(~\frac{\dot{x}^2}{4 D} + V_{eff}[x]~\right)},
\ee
where $\mathcal{N}$ is a normalization factor which comes from the Wiener measure and assures that $\int d x~ P(x, t|x_i) = 1$.
The expression (\ref{PI}) could have been obtained directly by computing the probability of path generated by iterating a discretized representation of the Langevin Eq. (\ref{Langevin}) --- see e.g. the discussion in \cite{elber1, RGMD}---. The advantage of the derivation given here is that it does not involve the stochastic calculus.

Eq. (\ref{PI}) provides a microscopic representation of the conditional probabilities, formulated in terms of the Langevin trajectories in configuration space  which connect 
$x_i$ and $x_f$. 
In the next sections we shall discuss how such conditional probabilities can be effectively evaluated using the DRP formalism.

The  path integral expression of the conditional probability current (\ref{Jdef}) is
\be
\label{JcondPI}
J(x_f, t|x_i) =  \frac{1}{2}~\left(  \langle v(x) \rangle_{x_i}^t -  D~\beta \nabla U(x)\right) ~P(x,t|x_i),
\ee
where $\langle v(x) \rangle_{x_i}^t$ denotes the average  velocity  the system reaches the configuration $x_i$ at time $t$, i.e.  
\be
\langle v(x) \rangle_{x_i}^t =  \frac{\int_{x_i}^{x} \mathcal D x~ \dot x(t)~e^{- S_{eff}[x]}}{\int_{x_i}^{x} \mathcal D x~ ~e^{- S_{eff}[x]}}.
\ee

\section{The DRP formalism}
\label{DRPform}

 
The DRP approach is based on the saddle-point approximation of the path integral  (\ref{PI}).
The expansion parameter controlling the accuracy of such an  approximation is the thermal energy $k_B T$, which enters in the definition of the diffusion constant $D$ and of the effective potential $V_{eff}(x)$. 
 
The saddle-points  are the paths with the highest statistical weight $\exp(-S_{eff}[x])$, i.e. those which  minimize the effective action functional
\be
 S_{eff}[x]= \int_{0}^t d\tau ~\left(~\frac{\dot{x}^2}{4 D} + V_{eff}[x]~\right).
 \ee
Note that all such so-called \emph{dominant reaction pathways} satisfy the boundary conditions 
\be
\label{BC}
\bar x(0) &=& x_i\nonumber\\
\bar x(t) &=& x_f.
\ee
By imposing the extremum condition $\delta S_{eff}[x]=0$, we obtain the equation of motion for the dominant reaction pathways $\bar x(t)$:
\be
\frac{1}{2D} \ddot{\bar x}(t) = \nabla V_{eff}[\bar x(t)].
\label{DRPeom}
\ee

In principle, a solution of the boundary-value problem (\ref{BC})-(\ref{DRPeom}) may be obtained by minimizing numerically a discretized representation of the effective action functional $S_{eff}[x]$.
In practice, however,  the presence of decoupling of time scales makes such a task very challenging.  Indeed, computing a single  dominant pathway would require to find a minimum of a function 
with a large number of degrees of freedom, $d \times N_t$, where $N_t$ the number of time discretization steps. 

Fortunately, a major numerical simplification of this problem can be achieved by exploiting the fact that the  equation of motion (\ref{DRPeom}) is simplectic, i.e. it conserves the "effective energy"
\be
E_{eff}= \frac{1}{4D} \dot{\bar x}^2(t) - V_{eff}[\bar x(t)].
\ee
Hence, rather than minimizing directly the effective action $S_{eff}[x]$, it is possible to obtain the dominant reaction pathways 
using the Hamilton-Jacobi (HJ) formulation of classical mechanics.  In other words, the trajectories obeying the equation of motion (\ref{DRPeom}) and subject to the boundary conditions (\ref{BC})
are those which minimize the effective HJ functional
\be
\label{SHJ}
S_{HJ}[x(l)]    = \frac{1}{\sqrt{D}}~\int_{x_i}^{x_f}d l \sqrt{E_{eff}(t) + V_{eff}[x(l)]},
\ee
where $dl = \sqrt{d x^2}$ is the measure of the distance covered by the system in configuration space,  during the transition. The advantage of the HJ formulation is that it allows to remove the time as an independent
variable. Instead, one introduces the curvilinear abscissa $l$. Since there is no gap in the length scales of molecular systems, the discretization of the HJ is expected to converge extremely much faster than the time discretization of the
effective action $S_{eff}[x]$. 

The effective energy $E_{eff}$ is an external parameter which determines the time at which each configuration of  a dominant reaction pathway is visited,  according to the usual HJ relationship
\be
\label{time}
t(x) =\int^{x}_{x_i} dl \frac{1}{ \sqrt{4 D(E_{eff}+V_{eff}[\bar x(l)])}}.
\ee
Typically, one is interested in studying transitions which terminate close the local minima of the potential energy $U(x)$. The residence time in such end-point configurations must be much longer than that in the configurations visited during the  transition. 
From Eq. (\ref{time}) it follows that these conditions are verified if
\be
E_{eff}\sim -V_{eff}(x_o).
\ee
where $x_o$ is a configuration in the vicinity of a local minimum of $U(x)$.

In practice, computing the dominant reaction pathway connecting two given configurations $x_i$ to $x_f$ amounts to minimizing a discretized version of the effective HJ functional:
\begin{equation} 
\label{effact_discr}
 S_{HJ}^{d} [x(l)]= \sum_{n=1}^{N_s-1} \sqrt{ \frac{1}{D} \left[
E_{eff} + V_{eff}\left( x(n) \right) \right] } \; \Delta l_{n,
n+1},
\end{equation}
where $N_s$ is the number of path discretization slices. Once such a path has been determined, one can reconstruct the time at which each of the configurations is visited during the transition, using Eq. (\ref{time}). In particular, the time interval between the $n$-th and the $(n+1)$-th slice is 
\be
\label{times}
\Delta t_{n+1, n} = \frac{\Delta l_{n+1, n}}{ \sqrt{4 D(E_{eff}+V_{eff}[\bar x(n)])}},
\ee
where $\Delta l_{n+1, n}= \sqrt{(x(n+1)-x(n))^2}$. For a discussion on how to efficiently perform the relaxation of the HJ action, we refer the reader to \cite{DRPappl1, DRPappl2}, where the DRP method is used to investigate \emph{ab-initio} chemical and conformational transitions in realistic molecular systems. 

Note that, in general, the solution of the boundary value problem (\ref{DRPeom})-(\ref{BC}) is not unique. Hence, one should  in principle take into account for the entire set of  dominant paths $\bar x_i(l)$ which obey the same boundary conditions (\ref{BC}). In practice, in many transitions of interest the relative statistical weight of secondary dominant paths with the same boundary conditions is much smaller and can be neglected. 

The dominant paths which solve the saddle-point equation (\ref{DRPeom}) can be used to estimate the time evolution of an arbitrary configuration-dependent observable $O(x)$, during a transition from the reactant to the product. 
To this end, let $h_R(x)$ and $h_P(x)$ be the  characteristic functions of the reactant and product states respectively ---i.e. $h_{R(P)}(x)=1$ if $x\in R(P)$ and $h_{R(P)}(x)=0$ otherwise--- and let $\rho_0(x)$ be the initial distribution of configurations in the reactant state. The average value of the observable $O(x)$ at some intermediate 
 time $0\le \tau\le t$, evaluated over all possible reactive pathways which visit the product  state at  time $t$ reads:
\be
\langle  O(\tau) \rangle &=& \frac{\int d x_f ~h_P(x_f) ~\int d x_i ~h_R(x_i)~ e^{-\frac{\beta}{2}(U(x_f) - U(x_i))}~\rho_0(x_i)~ \int_{x(0)=x_i}^{x(t)=x_f}~ \mathcal{D} x ~ O[x(\tau)] ~e^{-S_{eff}[x]}~ }
{\int d x_f ~h_P(x_f) ~\int d x_i ~h_R(x_i)~ e^{-\frac{\beta}{2}(U(x_f) - U(x_i))}~\rho_0(x_i)~ \int_{x(0)=x_i}^{x(t)=x_f}~ \mathcal{D} x ~e^{-S_{eff}[x]} }.
\ee

To lowest-order in the DRP saddle-point approximation (i.e. up to corrections of order $k_B T$) this average can be approximated with an average along the dominant paths only. In such 
an approximation, using the relationship
$S_{eff}[\ox] = - E_{eff} t + S_{HJ}[\ox]$, we have 
\be
\label{OLO}
\langle  O(\tau) \rangle &\simeq& \frac{\int d x_f ~h_P(x_f) ~\int d x_i ~h_R(x_i)~ e^{-\frac{\beta}{2}(U(x_f) - U(x_i))}~\rho_0(x_i)~ \sum_{k}~ ~ O[\bar x_k(\tau)] ~e^{E^k_{eff} t-S_{HJ}[\bar x_k]}~ }
{\int d x_f ~h_P(x_f) ~\int d x_i ~h_R(x_i)~ e^{-\frac{\beta}{2}(U(x_f) - U(x_i))}~\rho_0(x_i)~ \sum_{k}~e^{E^k_{eff} t-S_{HJ}[\bar x_k]} },
\ee
 where the sum $\sum_k$ runs over all the dominant paths $\bar x_k(t)$, with boundary condition $\bar x_k(0) = x_i$, $\bar x_k(t)=x_f$. In principle, the effective energy parameters 
 $E_{eff}^k$ have to be fixed in such a way that the total time of the transition is the same for all pathways. In practice,  the effects arising from  choosing the same value of $E_{eff}$ for all reaction pathways 
 are usually found to be negligibly small.  

Once  a dominant path $\bar x(\tau)$ has been determined, it is also possible to identify the configuration $x_{TS}$ along this path which belongs to the transition state (TS). 
This can be defined as the set of all configurations from which the system diffuses with probability  $1/2$ into the product, before visiting the reactant.
Clearly, the reactive pathways cross the transition state. In particular, the configurations of the dominant reaction paths which  are representative of the TS can be found by requiring  that the probability to 
diffuse back  to the initial  configuration $x_i$ along the saddle-point path,  equates that of evolving toward the final configuration.
 To the leading-order in the saddle-point approximation, this condition leads to the simple equation~\cite{DRP3}: 
 \be 
\label{transition} 
 \frac{U(x_f)-U(x_i)} {2 k_B T } =
\int_{x_{TS}}^{x_i} dl~\sqrt{\frac{1}{D}\left(E_{eff} + V_{eff}[\bar x(l)] \right)}-\int_{x_{TS}}^{x_f} dl~\sqrt{\frac{1}{D}\left(E_{eff} + V_{eff}[\bar x(l)] \right)}.  
\ee 
This equation can be easily solved for $x_{TS}$, once the dominant path $\ox(\tau)$ has been calculated. 

\section{Accounting for Small Fluctuations around the Dominant Paths}
\label{pra}

Let us now go beyond the lowest-order saddle-point approximation, and compute the contribution of the stochastic fluctuations around the dominant paths, to order $k_B T$. 
For sake of simplicity, in the following we shall  consider  the case in which the path integral associated to the conditional probability $P(x_f, t|x_i)$ 
has a single saddle-point. The generalization to multiple  dominant reaction pathways is straightforward: one simply needs to repeat such a calculation for each local maximum of the path probability. 

Let us consider  a completely general transition pathway $x(\tau)$ with boundary condition $x(t)=x_f$ and $x(0)=x_i$ and re-write it as the sum of the  dominant  trajectory $\bar x(\tau)$ and a fluctuation  $y(\tau)$ around it:
\be
x(\tau)= \ox(\tau)+ y(\tau).
\ee
Note that, by construction, the fluctuation path $y(\tau)$ satisfies the boundary conditions $y(0)=y(t)=0$.

We now provide an approximation to the path integral defining the conditional probability $P(x_f, t|x_i)$  by functionally expanding the action around  $\ox(\tau)$:
\be
\label{DRPexpansion}
S_{eff}[x] &=& S_{eff}[\ox]+ \frac{1}{2} \int_{0}^{t} d\tau' \int_{0}^{t} d\tau'' 
\frac{\delta^2 S_{eff}[\ox]}{\delta x_i(\tau') \delta x_k(\tau'')}~y_i(\tau)\,y_k(\tau) + \mathcal{O}(y^3)\qquad(i, k=1, \ldots d)\nonumber\\
&\simeq& S_{eff}[\ox] + \frac{1}{2} \int_{0}^{t} d\tau~\int_{0}^{t} d\tau' y_i(\tau) ~\hat{F}^{\tau, \tau'}_{i k}[\ox]~ y_k(\tau),
\ee
where we have introduced the so-called fluctuation operator $\hat{F}[\ox]$, defined as 
\be
\hat{F}^{\tau, \tau'}_{i k}[\ox] \equiv \frac{\delta^2 S_{eff}[\ox]}{\delta x_i(\tau') \delta x_k(\tau'')} =
\left[-\frac{1}{2D}~\delta_{i k }~\frac{d^2}{d \tau^2} + \partial_i \partial_k~V_{eff}[\ox(t)]\right] \delta(\tau-\tau').
\ee
Note that the fluctuation operator $\hat F[\ox]$ determines the amplitude of the  stochastic fluctuations around  the dominant path $\ox(\tau)$. We also stress that the expansion of the effective action does not contain linear terms in the fluctuation field $y(\tau)$, because the  dominant path $\bar x(\tau)$ around which we expand is a solution of the equations of motion (\ref{DRPeom}), i.e. a stationary point
of the effective action functional, $S_{eff}[x]$. 

The functional integral over the fluctuation field $y(\tau)$ can be performed formally. To this end, expand the fluctuation function $y(\tau)$ in the
 basis of the (real) complete set of eigenfunctions of $\hat{F}[\ox]$:
\be
y(\tau)= \sum_n c_n x_n(\tau), \qquad
\hat{F}[\ox]~ x_n(\tau) = \lambda_n ~x_n(\tau).
\ee
where the eigenfunctions $x_n$ are  chosen so as to satisfy the boundary conditions
$x_n(0)=x_n(t)=0$ and the normalization condition 
\be
\int_{0}^{t} d\tau ~x_m(\tau)~ x_n(\tau) = \delta_{m n} 
\ee
To second order in the fluctuations, the action reads
\be
S_{eff}[x] = S_{eff}[\ox]+ \frac{1}{2}~\sum_n~ c^2_n ~\lambda_n + ...
\ee
and the measure of the functional integral can be re-written as
\be
\mathcal{D} x = \mathcal{D} y =  \prod_n~ \int_{-\infty}^{\infty} \frac{d c_n}{ \sqrt{(2\pi)^{d}}}.
\ee
Hence, the conditional probability  can be written as
\be
\label{KtWKB}
P(x_f,t|x_i)&\simeq&  e^{-\frac \beta 2 (U(x_f)-U(x_i))} ~\mathcal{N}~\int_{x(0)=x_i}^{x(t)=x_f} \mathcal{D} x~e^{-\int_{0}^t d\tau ~\left(~\frac{\dot{x}^2}{4 D} + V_{eff}[x]~\right)}\\
&\simeq &\mathcal{N}~e^{-\frac{\beta}{2}(U(x_f)-U(x_i))}~e^{-S_{eff}[\ox]}\int_{-\infty}^{\infty}~\prod_n \frac{d c_n}{ \sqrt{(2\pi)^d}}
e^{- \lambda_n~ c_n^2}  \nonumber \\
&=&  \mathcal{N}~\frac{e^{-\frac{\beta}{2 }(U(x_f)-U(x_i))}}{\sqrt{\det \hat{F}[\ox]}}~e^{-S_{eff}[\ox]}.
\ee

Eq. (\ref{KtWKB}) does not yet provide a practical tool  to compute the conditional probability, since both the normalization factor $\mathcal{N}$  and the determinant of the fluctuation operator $\det \hat F[\ox]$  diverge in the continuum limit. 
However, there exist a standard trick to represent the ratio $\frac{\mathcal{N}}{\sqrt{\det F}}$ in a form which is finite and calculable in the continuum limit. 
The idea is to introduce a so-called \emph{regulator}, i.e. to multiply and divide by the conditional probability density  of a fictitious system $P_{reg.}(x_f, t|x_i)$. Such as system must be chosen in such a way that (i) 
the solutions of the Smoluchowski equation are known analytically and (ii) the DRP expression (\ref{KtWKB}) to second-order in the fluctuations provides the complete exact result. 
 
For example, one can use as regulator the conditional probability associated to the diffusion in an external \emph{harmonic} potential such as 
\be\label{UHO}
U_{HO}(x) \equiv~\frac{1}{2}~\alpha~(x-x_0)^2.
\ee
The solution of the Smoluchowski equation for such a simple system are known analytically. In particular, in appendix \ref{PHOapp} we show that for $x_f=x_i=x_0$ and choosing the parameter $\alpha$ in such a way that
 $\alpha \,\beta\, D\, t\,\gg1$  one has:
 \be
P_{reg.}(x_0, t|x_0) &\simeq&  \left(\frac{\beta \alpha}{2 \pi}\right)^{d/2}.
 \ee
It is immediate to verify that, for such a system, all the contributions to the expansion (\ref{DRPexpansion}) beyond the second order in the fluctuation field $y(\tau)$ vanish identically. Hence, the second-order DRP approximation yields in fact the correct exact result. 
Alternatively, one may use as regulator the  conditional probability associated to the free Brownian motion:
\be
P_{reg.}(x_0, t|x_0) =  \left(\frac{1}{4 \pi D t}\right)^{d/2}.
\ee
Also for such a system, there is no contribution to the DRP expansion beyond the second order.

In order to remove the divergences in  Eq. (\ref{KtWKB}),  we multiply and divide by  $P_{reg.}(x_0, t|x_0)$ 
and we replace the term in the denominator with its (exact) second-order DRP saddle-point representation:  
\be
P_{DRP}(x_f,t|x_i)&=& \mathcal{N}~\frac{e^{-\frac{\beta}{2 }(U(x_f)-U(x_i))}}{\sqrt{\det \hat{F}[\ox]}}~e^{-S_{eff}[\ox]}\nonumber\\\
&=& \frac{ P_{reg.}(x_0, t|x_0)}{P_{reg.}(x_0,t|x_0)}~\mathcal{N}~\frac{e^{-\frac{\beta}{2 }(U(x_f)-U(x_i))}}{\sqrt{\det \hat{F}[\ox]}}~e^{-S_{eff}[\ox]} 
= \frac{ P_{reg.}(x_0, t|x_0)}{\mathcal{N}~\frac{1}{\sqrt{\det \hat{F}_{reg.}[\ox_{reg.}]}}~e^{-S^{reg.}_{eff}[\bar x_{reg.}]}}\mathcal{N}~\frac{e^{-\frac{\beta}{2 }(U(x_f)-U(x_i))}}{\sqrt{\det \hat{F}[\ox]}}~e^{-S_{eff}[\ox]}\nonumber\\
&=& P_{reg.}(x_0, t|x_0)~e^{-\frac{\beta}{2 }(U(x_f)-U(x_i))}~e^{S^{reg.}_{eff}[\ox_{reg.}]-S_{eff}[\ox]}~\sqrt{\frac{1}{\det \left(\hat{F}^{-1}_{reg.}[\ox_{reg.}]~\hat{F}[\ox]\right)}} \nonumber\\\
&=&  P_{reg.}(x_0, t|x_0)~e^{-E^{reg.}_{eff} t + S_{HJ}^{reg.}[\ox_{reg.}]} ~\exp\left[-\frac{\beta}{2 }(U(x_f)-U(x_i)) + E_{eff} t[\ox] - S_{HJ}[\ox] -\frac{1}{2}~\textrm{Tr}\log \left( \hat{F}^{-1}_{reg.}[\ox_{reg.}]~\hat{F}[\ox]\right) \right], \nonumber\\
\label{main}
\ee
where $E_{eff}^{reg.}$ is the value of the effective energy parameter for which the total time in the regulator conditional probability is the same as in the probability $P_{DRP}(x_f, t| x_i)$ we want to compute, i.e.
\be
t[\ox] = \int_{x_i}^{x_f} \frac{dl}{\sqrt{4 D (E_{eff}+ V_{eff}[\ox])}} = \int_{x_i}^{x_f} \frac{dl}{\sqrt{4 D (E^{reg.}_{eff}+ V^{reg.}_{eff}[\ox_{reg.}])}}= t_{reg.}[\ox_{reg.}]
\ee

Some comment on the expression (\ref{main}) are in order. 
First of all, we observe that the factor $\det \left(\hat{F}^{-1}_{reg.}[\bar x_{reg.}]~\hat{F}[\ox]\right)$ remains finite  in the continuum limit. 
Then we note that the $o(k_B T)$ correction in the DRP formalism is the analog of the Ginzburg correction of statistical field theory, and of the one-loop correction 
 of quantum field theory.  Finally, we emphasize again the fact that the regulator does not need to have any physical interpretation. It has been introduced as a mere mathematical trick to regulate the divergences appearing in Eq. (\ref{KtWKB}). 

From Eq. (\ref{main}) it is straightforward to obtain the DRP expression for the probability current,
\be
J(x_f, t| x_i) &=& -D \left. (\nabla  + \beta \nabla U(x) ) P(x,t|x_i)\right|_{x=x_f}\nonumber\\
&=& D \left( \sqrt{\frac{E_{eff}+V_{eff}(x_f)}{D}} \hat u_\theta(x_f) -\frac{\beta}{2} \nabla U(x_f) \right) P_{DRP}(x_f, t|x_i)
\label{DRPcurrent}
\ee
where $\hat u_\theta(x)$ is the versor tangent to the dominant path at the configuration $x$.

We note that in deriving Eq. (\ref{DRPcurrent}), we have neglected the contribution coming form the gradient of the fluctuation determinant, since these term provides corrections which are of higher 
order in $k_B T$.  The $o(k_B T)$ DRP expression for the probability current will be used in section~\ref{rates} to compute the reaction rates.

The DRP formula for the time evolution of the average of the observable $O(x)$ including $o(k_B T)$ corrections reads:
\be
\label{ONLO}
\langle  O(\tau) \rangle &\simeq& \frac{\int d x_f ~h_P(x_f) ~\int d x_i ~h_R(x_i)~ e^{-\frac{\beta}{2}(U(x_f) - U(x_i))}~\rho_0(x_i)~ \sum_{k}~ ~ O[\bar x_k(\tau)] ~e^{E^k_{eff} t-S_{HJ}[\bar x_k]}
\sqrt{\det \left( \hat{F}_{reg.}[x_0]~\hat{F}^{-1}[\ox_k]\right) }}
{\int d x_f ~h_P(x_f) ~\int d x_i ~h_R(x_i)~ e^{-\frac{\beta}{2}(U(x_f) - U(x_i))}~\rho_0(x_i)~ \sum_{k}~e^{E^k_{eff} t-S_{HJ}[\bar x_k]} \sqrt{\det \left( \hat{F}_{reg.}[x_0]~\hat{F}^{-1}[\ox_k]\right) }}.
\ee
As in Eq. (\ref{OLO}),  the sum $\sum_k$ runs over all the dominant paths $\bar x_k(t)$.

In practice, the  determinant $\det \left(  \hat{F}_{reg.}[\ox_{reg.}]~\hat{F}^{-1}[\ox]\right)$ has to be evaluated numerically from 
 a discretized representation of the fluctuation operator associated to  the dominant path $\hat{F}[\ox]$. To obtain such a representation,   one needs to express the time derivative using discretized time intervals.
 It is most convenient to use the intervals $\Delta t_{i, i+1}$ evaluated from the dominant trajectory, according to Eq. (\ref{times}). The fluctuation operator reads
\be
\label{discrF}
\hat{F}[\bar{x}]^{i,j}_{k,m} = \frac{-1/D}{\mmp+\mmn} \delta_{i,j} \, \left[\frac{\delta_{k,m+1}}{\mmp}- \delta_{k,m} 
      \left( \frac{1}{\mmp}+ \frac{1}{\mmn} \right) +\frac{\delta_{k,m-1}}{\mmn}  \right] &+
    \frac{\partial^{2}\veff(\bar{x}(k))}{\partial x_{i} \partial x_{j}} \delta_{k,m},
    \ee
    where the indexes $k, m=1, \dots, N_s$ run over the path frames, while the indexes $i, j=1, \ldots, d$ label the degrees of freedom of the system. 
\begin{figure}[t]
\includegraphics[width=8.5 cm]{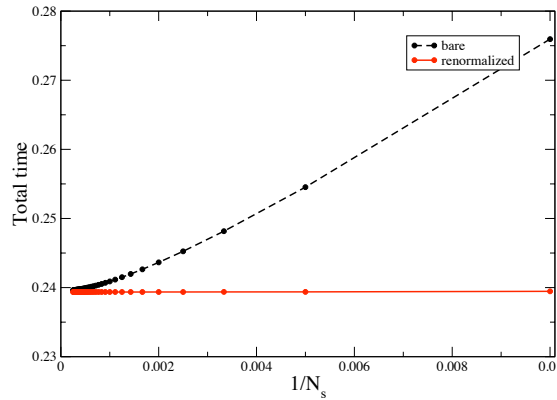}\qquad
\caption{The total time for a transition from $x_i=-1$ to $x_f=1$ in the quartic potential (\ref{Umod1}), calculated using Eq. (\ref{times}) with different number of discretization steps, in 
the original theory and in the EST. }
\label{doublewell}
\end{figure}

\section{Improving the Convergence on the DRP Calculation}
\label{convergence}

The main numerical advantage of the DRP formalism arises from the possibility of replacing the time as a dynamical variable, and replace it with the curvilinear abscissa $l$. Since there is no gap in 
the length scales of molecular systems, the convergence of the discretization of $l$ is usually very fast. As a result, using such a formulation, it
is possible to gain  information about the reaction mechanism at a very low computational cost. 
 
On the other hand, in order to obtain information about the dynamics, one needs to compute the times at which each configuration is visited along the dominant path, using Eq.s (\ref{time}) and (\ref{times}).  
The calculation of the time intervals $\Delta t_{i,i+1}$ is also needed in the discretized representation of the fluctuation operator (\ref{discrF}), which enters in the 
calculation of the order $k_B T$ corrections arising from non-equilibrium stochastic fluctuations around the dominant path. 
 
A potential limitation of the DRP approach resides in the fact that the DRP calculation of the time intervals converges very slowly with the number of equal displacement discretization steps. 
As a result, in order to achieve an accurate description of the dynamics, or in order properly take into account  of the effects of fluctuations, one needs to use a large number of path 
frames, with a consequent significant  increase of the computational cost of a DRP simulation. 

To illustrate this problem in a simple example, let us consider the diffusion of a particle in a one-dimensional quartic external potential
\be
\label{Umod1}
U(x) =  \alpha (x_0-x^2)^2.
\ee
We shall adopts units in which $x_0=1$, $\beta=1/k_B T=5$ and $D=1$. 

The dashed line in the left panel of Fig.\ref{doublewell} shows the result of the calculation of the total time interval for a dominant transition from $x_i=-1$ to $x_f=1$, as a function of 
the inverse of the number of equally-displaced path discretization steps $N_s$. Such a time interval was computed by adding up all the elementary time intervals $\Delta t_{i, i+1}$ 
evaluated according to Eq. (\ref{times}), by choosing  $E_{eff}=-V_{eff}(x_i)+ 0.03$. This figure shows that, in order to reduce the discretization errors below $1 \%$,
 one needs to use  $\sim 10^3$ path discretization steps. 

Such a slow convergence is of course a consequence of the decoupling of the time scales characterizing the dynamics of this system. Indeed, the quasi-free 
diffusion in the bottom of the wells is much slower than the crossing of the transition regions, where the force is large. An analog convergence problem
is  encountered also in molecular systems, since the characteristic internal time scales are decoupled and range from fractions of ps to ns.  

The purpose of this section is to show that the convergence of the calculation of the time intervals in the DRP approach can be greatly improved by adopting the effective stochastic theory (EST) 
developed in \cite{EST} and briefly reviewed in the appendix \ref{ESTapp}.  

The main idea of the EST is to exploit the gap in the internal time scales in order to \emph{analytically} perform the integral over  the fast Fourier components of the  
paths $x(\tau)$ which contribute to the path integral (\ref{PI}). Through such a procedure, the effects of the fast dynamics is rigorously and systematically averaged out
by "renormalizing"  the effective potential:
\be
\label{VR1}
V_{eff}(x) &\rightarrow& V^{EST}_{eff}(x) = V_{eff}(x) + V^{R}_{eff}(x).\\
\label{VR2}
V^{R}_{eff}(x) &=& \frac{D \Delta t_c~(1-b)}{2 \pi^2 b} \nabla^2 V_{eff}(x) + \ldots
\ee 
In such an Eq., $\Delta t_c$ is a cut-off time scale which must be chosen much smaller than the fastest internal dynamical time scale and $b$ is a parameter which defines the interval of Fourier modes which are
 being analytically integrated out --- see the discussion in the appendix \ref{ESTapp}---. Typically, for molecular systems  $\Delta t_c \sim 10^{-3}$ps and $b\sim 10^{-2}$ ~\cite{RGMD}.
 The dots in Eq. (\ref{VR2}) denote higher order correction in an expansion in the ratio of slow and fast time scales (slow-mode perturbation theory).

The EST generates by construction the same long-time dynamics of the original --- or so-called "bare"--- theory, but has a lower time resolution. In the context of MD simulations, this implies that the EST
 can be integrated using much larger discretization time steps~\cite{RGMD}. 
In the  context of the  DRP simulations, the utility of the EST resides in the fact that fewer path discretization time steps are required in order to achieve a convergent calculation 
of the time interval from the dominant path, through Eq. (\ref{time}). 
Such a gain is clearly visible in  Fig. \ref{doublewell}, where we compare the total time interval obtained in the bare theory and in the EST, for different numbers of path discretization steps $N_s$. 
From the mathematical point of view, the computational gain of adopting the EST  can be seen as a consequence of the fact that the renormalized effective potential $V_{eff}^{EST}(x)$ is in general a smoother
 function than the bare effective potential $V_{eff}(x)$.

 \section{Testing the DRP calculation with  $o(k_B T)$ corrections on an analytically solvable model}
 \label{test1}
 
In this section, we assess the accuracy of the DRP $o(k_B T)$ calculation developed in the previous section 
 by computing the conditional probability and the probability current in an exactly solvable model. In particular, let us consider a one-dimensional point particle diffusing in a harmonic oscillator
 of potential
 \be
 U_{HO} = \frac{1}{2} \alpha~ x^2.
 \ee 
 
 The conditional probability for the point particle  to  be at the origin  $x_i=0$ at the initial time and to reach the point $x_f$ after a time interval $t$ is known analytically and reads:
 \be
 \label{exactHO}
 P_{HO}(x_f,t|x_i=0) = \sqrt{\frac{1}{4 \pi}~\frac{\alpha \beta}{\sinh (\alpha~\beta~D~t) }} 
 ~\exp\left\{-\frac{\alpha~\beta~x_f^2~\cosh(\alpha~\beta~D~t)}{4  \sinh(\alpha ~\beta~D~t)} + \frac{1}{2}\alpha~\beta~D~t\right\}
 \ee
 
Let us now discuss  the DRP calculation of the same conditional probability. 
We choose to use as regulator the conditional probability of the free Browian diffusion  for $x_f=x_i$, i.e.  
\be
  P_{0}(x_i,t|x_i) = \sqrt{\frac{1}{4 \pi D t}}
  \ee
The NLO  DRP result is therefore
  \be
  \label{DRPHO}
  P_{DRP}(x_f,t|x_0) = e^{-\frac{\beta}{4}\alpha^2~\left(x_f^2-x_0^2\right)} \sqrt{\frac{1}{4 \pi D t}} ~\sqrt{\frac{\det F_{0}[x_i]}{\det F_{HO}[\bar x]}} e^{E_{eff} t[\bar x]-S_{HJ}[\bar x]}, 
\ee
where $F_{HO}[\bar x]$ is the fluctuation operator of the harmonic oscillator evaluated along the dominant path $\bar x(\tau)$ evaluated numerically, $F_0[x_i]$ is the fluctuation operator 
of the free Brownian theory, evaluated on the 
static path $x(t)=x_i$ --- with $S_{eff}[x_i]=0$---, while $t[\bar x]$ and
$S_{HJ}[\bar x]$ are computed from the dominant path  $\bar x(\tau)$ using Eq.s (\ref{SHJ}) and (\ref{time}), respectively.

\begin{figure}[t]
\includegraphics[width=8.5 cm]{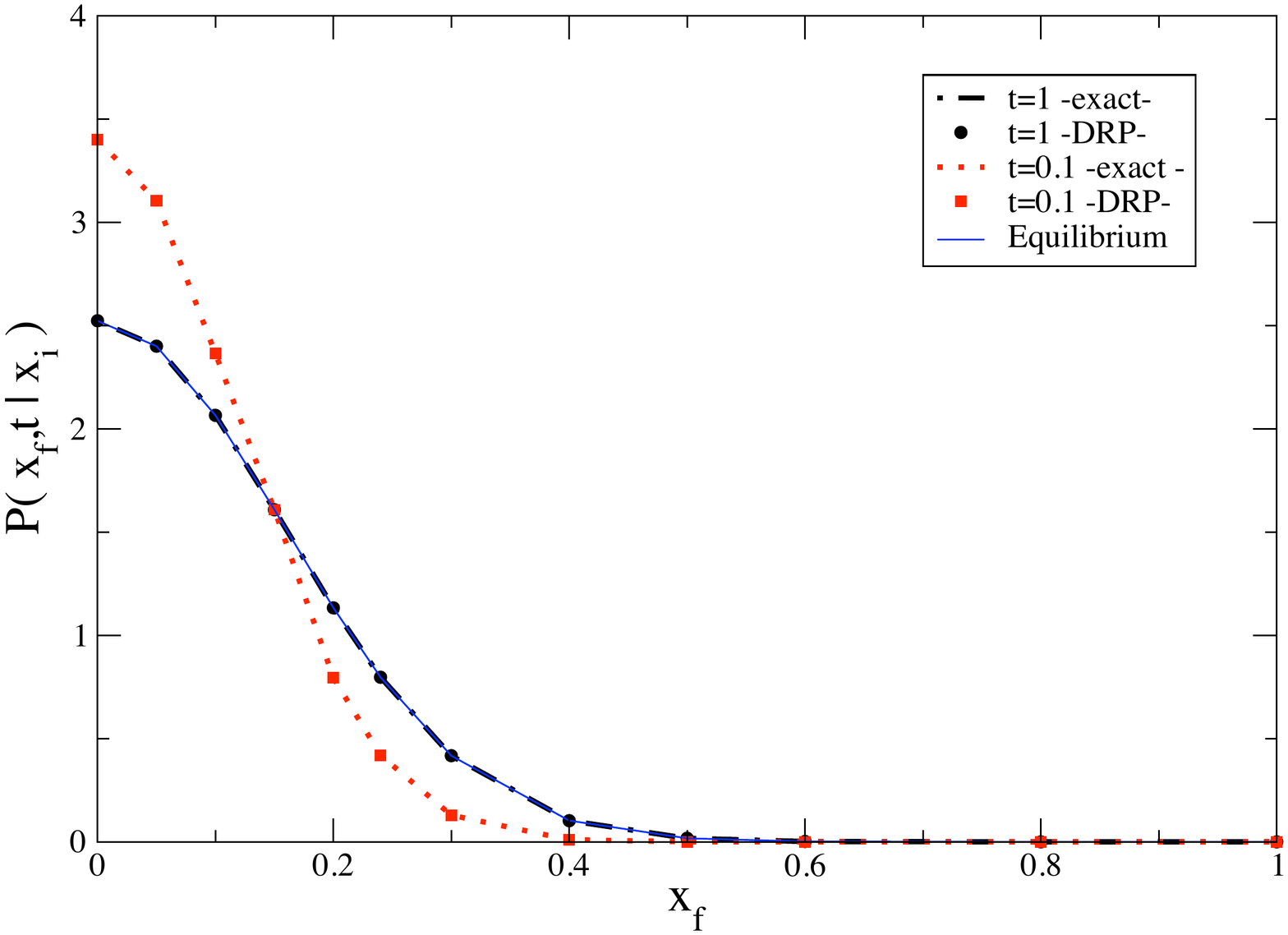}\qquad
\includegraphics[width=8.5 cm]{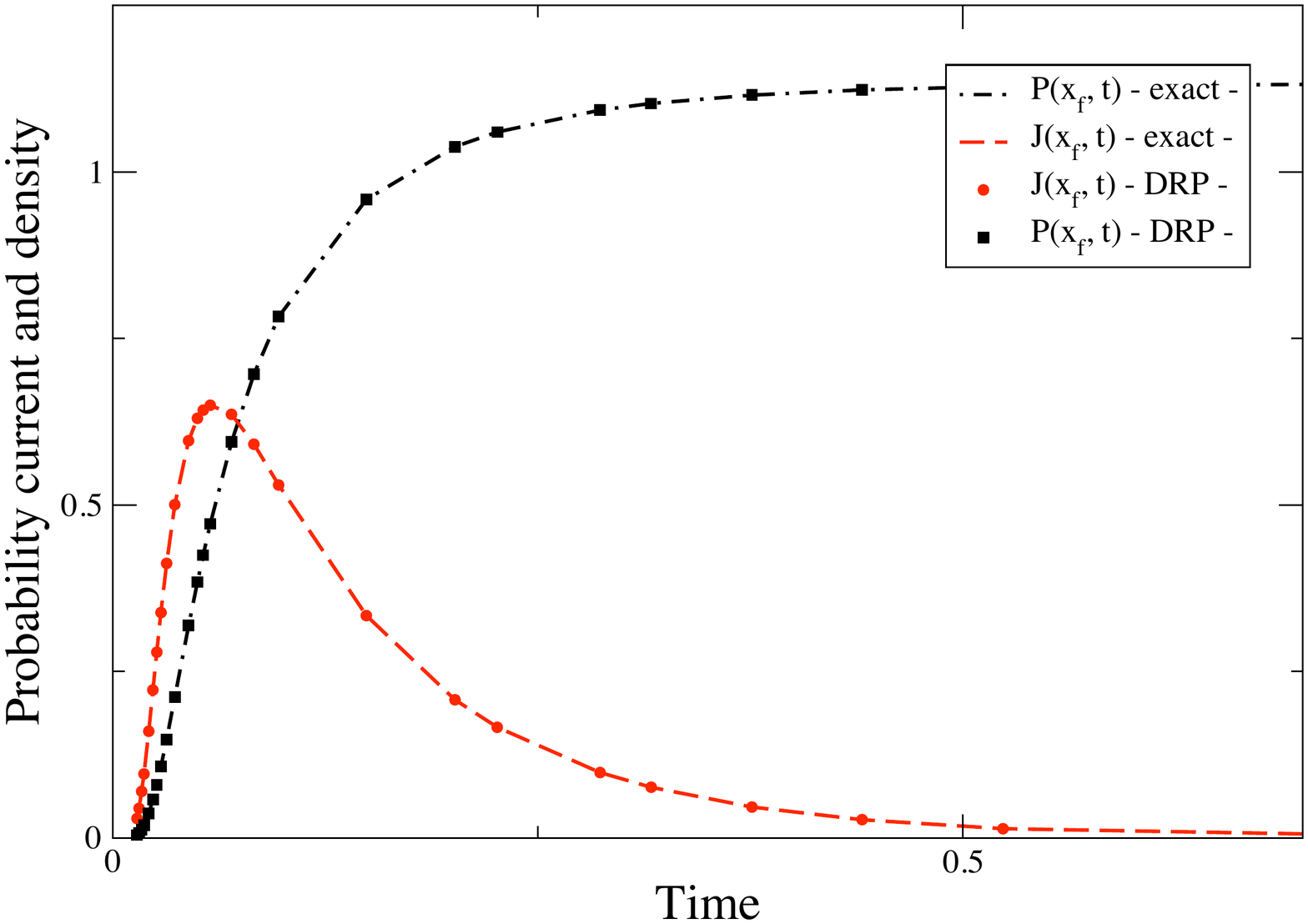}\qquad
\caption{Comparison between the DRP expression for the conditional probability for the diffusion in a harmonic potential, in a system of units in which $\alpha=4$  $\beta=10$ 
and $D \beta =1.$
In the left panel 
we show the DRP and the corresponding exact conditional probability $P(x_f, t|x_i)$   as  function of the final position  $x_f$, for $x_i=0$ and for two times $t=0.1$ and $t=1$. We also show the corresponding Boltzmann equilibrium distribution. 
In the right panel, we compare the DRP prediction 
for the probability density and for the probability current with the corresponding exact results, as a function of the time interval $t$, at a fixed final position $x_f=0.2$.  }
\label{exactvsDRP}
\end{figure}

We recall that, in the specific case of the diffusion in an harmonic oscillator,  all the contributions to the DRP saddle-point expansion beyond the second order vanish identically.
 Hence, the NLO prediction (\ref{DRPHO}) \emph{must agree} to numerical accuracy with the analytic result (\ref{exactHO}), for all choices of $t$ and  $x_f$. 
 In the left panel of Fig. \ref{exactvsDRP} we compare the DRP and the exact analytic conditional probability for $x_i=0$  as  function of the final position  $x_f$, at the fixed times $t=0.1$ and $t=1$ for $\beta =10$. We note 
that at  long times --- i.e. $t=1$ ---- the system has attained thermal equilibrium (i.e. the Boltzmann distribution). In addition, the DRP calculation yields the exact result at any intermediate time, as expected.  
In the right panel of Fig. \ref{exactvsDRP} 
we compare the DRP prediction for the probability density and for the probability current with the corresponding exact results, as a function of the time interval $t$, 
at a fixed final position $x_f=0.2$ for $\beta =10$.  
We see that, in the long time limit, the probability current
progressively dies out as the corresponding probability distribution 
becomes stationary. As in the previous case,  the result of the numerical DRP calculation completely agrees with the expected analytic prediction. 
 
\section{Reaction Kinetics}
\label{rates}

In the previous sections we have discussed and tested the numerical evaluation of the conditional probability $P(x_f, t|x_i)$ and of the probability current $J(x_f, t|x_i)$, in   the DRP approach. 
 In this section, we show how such quantities can be used to microscopically compute the reaction rate.
 
Let us consider the case of a conformational reaction involving two thermodynamically (meta)-stable states, which we shall refer to as to the reactant $R$ and product state $P$, respectively.
The thermodynamical states are usually defined in terms of an order parameter  $O$,  i.e. a quantity which is distributed around 
different values in the reactant  state $R$ and in the product state $P$,  i.e. $O \simeq O_R$ in $R$ and  $O\simeq O_P$ in $P$. 
Hence, along the reaction pathways,  $O$ varies very rapidly from $O_R$ to $O_P$.

We shall further assume that the time scales in which the thermalization is achieved in each of the two-states is much smaller than the inverse rate of transitions between them.
 Under such conditions, the system is said to obey two-state kinetics.  In this case, the  fraction of population in the reactant and product obey the well-known kinetic Eq.s:
\be
\label{rkin1}
\dot n_R(t) &=& - k_{R\to P} ~n_R(t) +~ k_{P\to R}~ n_P(t)\\
\label{rkin2}
\dot n_P(t) &=& ~\ k_{R\to P}~ n_R(t)  - ~k_{P\to R} ~ n_P(t).
\ee
with the condition $n_R(t) +n_P(t)=1$.  

At  equilibrium, the population fraction stops depending on time, $n_{R(P)}(t)= n^{eq}_{R(P)}$, and Eq.s (\ref{rkin1})-(\ref{rkin2}) yield  the detailed balance condition,
\be
\frac{n_R^{eq}}{n_P^{eq}}=\frac{k_{P\to R}}{k_{R\to P}}.
\ee

The solution of the kinetic Eq.s (\ref{rkin1})-(\ref{rkin2}) with boundary conditions $n_P(0)=0, n_R(0)=1$ ---i.e. assuming that the  system is initially prepared in the reactant state--- are
\be
\label{solrkin1}
n_P(t) &=& n_P^{eq} \left(1- e^{-k t} \right),\\
\label{solrkin2}
n_R(t) &=& 1- n_P(t),
\ee
where $k = k_{R\to P}+ k_{P\to R}$ is  the thermal relaxation rate.

In the opposite short-time regime, i.e. for  $t \ll \frac{1}{k}$,  $n_P(t)$ grows linearly with the reaction rate $k_{R\to P}$:
\be
\label{NPshort}
n_P(t) \simeq k_{R\to P}~ t \qquad (k_{R\to P}~ t \ll 1).
\ee
Hence, the reaction rate $k_{R\to P}$ can be written as
\be
\label{shorttK}
k_{R\to P}  = \lim_{t\to0}\frac{d}{dt} n_P(t). 
\ee
The limit in such an equation means that the time $t$ is chosen much smaller than the inverse reaction rate --- i.e. $ k_{R\to P}~ t \ll 1$---,  yet much larger than all the thermalization time scales in the 
reactant state. The possibility of making such a choice is guaranteed by the assumption of two-state kinetics.
 
 In order to microscopically compute the rate in the DRP approach, we need the path integral expression of the population fraction $n_P(t)$, solution of the Eq.s (\ref{rkin1})-(\ref{rkin2}) with boundary 
 condition $n_P(0)=0$.  In the appendix \ref{PInP} we show that such a solution is given by
 \be
 \label{nPPI}
 n_P(t) &=&  \int d x_f \ h_P(x_f) \int d x_i \ h_P(x_i) \ \rho_0(x_i)~P(x_f,t|x_i)\nonumber\\
&=& ~\mathcal{N}~\int d x_f \ h_P(x_f) \int d x_i \ h_P(x_i) \rho_0(x_i)~ e^{-\frac \beta 2 (U(x_f)-U(x_i))} ~\int_{x(t_i)=x_i}^{x(t)=x_f} \mathcal{D} x~e^{-\int_{0}^t d\tau 
\left(\frac{\dot{x}^2}{4 D} + V_{eff}[x]\right)},
 \ee
where  $\rho_0(x)$ is the initial normalized distribution of the configurations in the reactant state and $h_R(x)$ and $h_P(x)$ are the characteristic functions of the reactant and product state, respectively. 

Eq. (\ref{shorttK}) becomes
\be
\label{shorttimePI}
k_{R\to P} &\stackrel{(k_{R\to P} t \ll 1)}{=}&  \int d x_f \ h_P(x_f) \int d x_i \ h_P(x_i) \rho_0(x_i)~\frac{\partial}{\partial t} P(x_f,t|x_i).
\ee
Using the Smoluchowski Eq. we can rewrite this as
\be
k_{R\to P} &\stackrel{(k t \ll 1)}{= }& \frac{d}{dt} n_P(t) = \int d x_f h_P(x_f) \int d x_i h_R(x_i) \rho_0(x_i) \frac{\partial}{\partial t} P(x_f, t|x_i)\\
&=&  \int  d x_f h_P(x_f) \int d x_i h_R(x_i) \rho_0(x_i) \nabla \cdot J(x_f, t|x_i)
\ee
We now introduce a closed surface $\partial W$ which surrounds the product state. Using Gauss's divergence theorem, we find: 
\be\label{int1}
k_{R\to P} &\stackrel{(k t \ll 1)}{= }&  \int d x_i~ h_R(x_i) ~\rho_0(x_i)~ \int_{\partial W} ~ d\sigma   \cdot J(x_f, t|x_i)\\
\label{int2}
&=&  - \int_{\partial W} ~ d|\sigma|~ \hat n_x   \cdot \overline{J(x_f, t|x_i)}
\ee 
where $\overline{J(x_f, t|x_i)}$ denotes the average of the current with respect to the initial configuration in the reactant, i.e. 
\be
\overline{J(x_f, t|x_i)} \equiv \int d x_i~ h_R(x_i) ~\rho_0(x_i)~ J(x_f, t|x_i),
\ee
and $\hat n_x$ is the unitary vector normal to the dividing surface $\partial W$ at the point $x$, oriented out-ward. 
Note that if the surface $\partial W$ is chosen in such a way that it intersects the transition region, then  Eq. (\ref{int2}) yields in fact a multi-dimensional generalization of Kramers' flux-over-population expression for the reaction rate\cite{review}.

The Eq.(\ref{int1}) contains the integrals over two large dimensional spaces, which may be difficult to perform in practical applications. Hence, it is useful to introduce further approximations.
First of all, we observe that under the assumption of two-state kinetics, the current $J(x_f, t|x_i)$ becomes quasi-instantaneously 
independent on the initial configuration $x_i$. This fact allows to remove the average over the initial configurations in the reactant, since for any choice of $x_i\in R$ one has 
\be
\overline{J(x_f, t|x_i)} \simeq J(x_f, t|x_i) \qquad \forall x_i\in R. 
\ee
Let us consider first the case in which  the reaction occurs through a single channel, i.e. that
all the stochastic paths starting form $x_i$ and ending up in the product after a short time $t$ are confined in a small bundle around a single dominant reaction pathway.  
In order to estimate the flux of the current through the dividing surface, we insert the DRP expression for the current, i.e. 
\be
\label{inter}
k_{R\to P} &\simeq& - \int_{\partial W} ~ d|\sigma|_x~ \hat n_x   \cdot J_{DRP}(x, t|x_i) \simeq
  P_{reg.}(x_0, t|x_0)~e^{-E^{reg.}_{eff} t + S_{HJ}^{reg.}[x_0]} \nonumber\\
  &\times& \int_{\partial W} ~ d|\sigma|_x~ \hat n_x   \cdot \frac{1}{2}\left(\langle \dot x(x) \rangle_{x_i}^t - \beta
\nabla U(x) \right)~e^{-\frac{\beta}{2} (U(x)-U(x_i))}~e^{E_{eff} t -S_{HJ}[\ox]} \sqrt{\det(F_{reg}[x_0] F^{-1}[\ox ])}
\ee
\begin{figure}[t]
\includegraphics[width=7 cm]{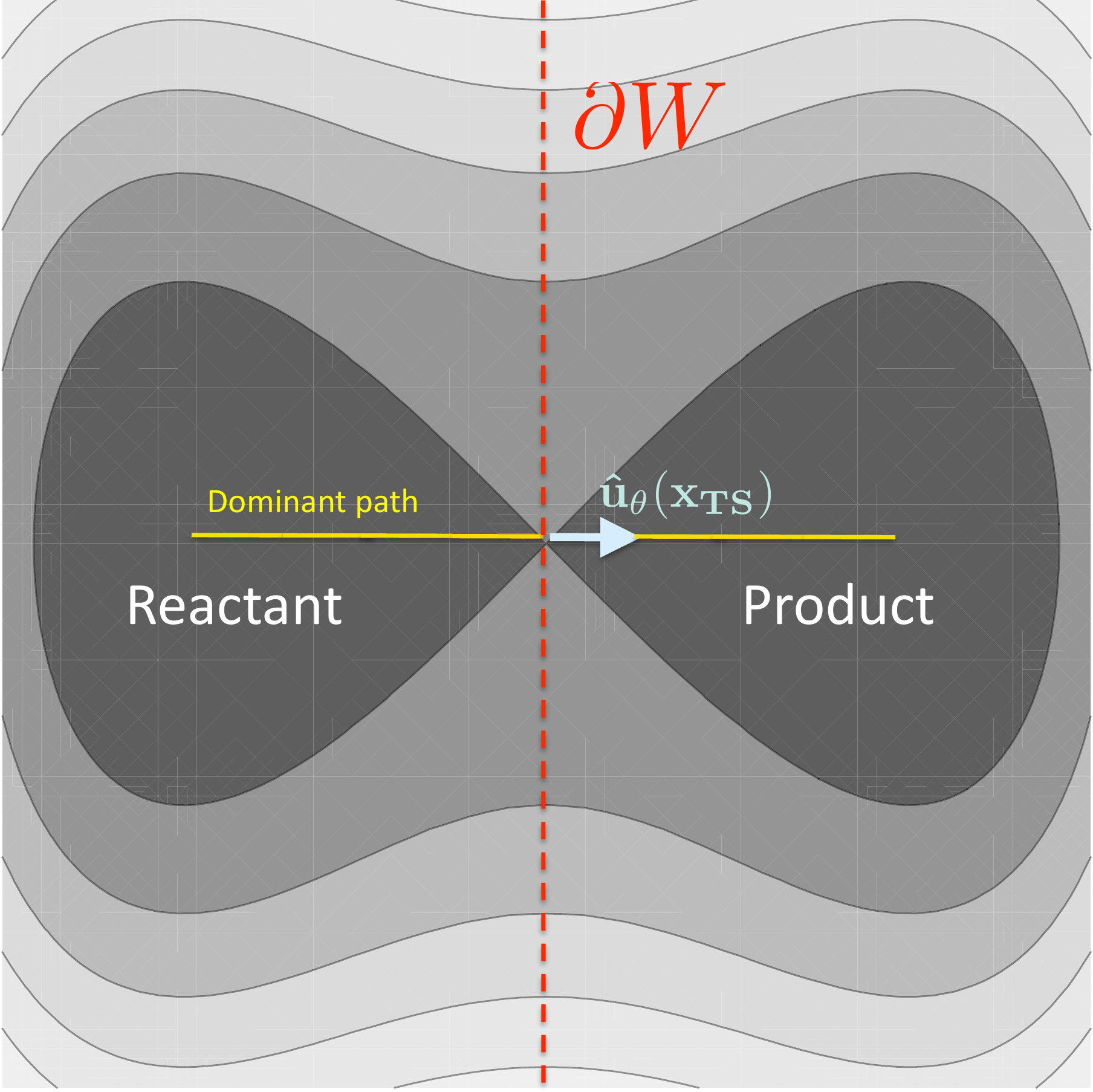}\qquad
\caption{The definition of  $\partial W$ as the hyper-surface orthogonal to the dominant path the the point $x_{TS}$ solution of the transition state Eq. (\ref{transition}).  }
\label{TSFIG}
\end{figure}

In the low-temperature limit, the leading dependence on $x$ in the integrand of Eq. (\ref{inter}) comes from the exponential factors and we can neglect the dependence on $x$ of all 
the non-exponential terms. In addition, in the same limit, we can use the approximation
\be
 S_{HJ}[\ox]  \simeq \frac{1}{2 k_B T} \int_{x_i}^x dl~ |\nabla U[\ox(l)]| = \frac{\beta}{2} (U(x)-U(x_i))
\ee
Hence, up to corrections of higher order in $k_B T$,  the flux (\ref{inter}) takes the  form  
\be
\label{general}
k_{R\to P} = - \int_{x\in  \partial W} ~ d|\sigma|_x~ \hat n_x   \cdot J(x, t|x_i) \simeq - Z_{\partial W}~e^{\beta U(x_{TS})}~ (\hat n_{x_{\partial W}}~ \cdot J_{DRP}(x_{\partial W},t|x_i)),
\ee
where $x_{\partial W}$ is a configuration on the dividing surface $\partial W$ and $Z_{\partial W}$  is the partition function
\be
\label{ZpartialW}
Z_{\partial W} =  \int_{x\in  \partial W} ~ d|\sigma|_x~  e^{-\beta U(x)}.
\ee

Eq. (\ref{general}) is  independent on  the specific choice of the dividing surface.  
We now specialize to the case in which  $\partial W$ is  the hyper-surface orthogonal to the dominant reaction pathway, at the point $x_{TS}$  solution of the transition state Eq. (\ref{transition})
 --- see Fig.~\ref{TSFIG}~---. 
With such a choice, 
\be
\hat n_x &=& - \hat u_\theta(x_{TS}),\qquad
\hat u_\theta(x_{TS}) \equiv \frac{\dot \ox(l_{TS})}{|\dot  \ox(l_{TS})|}.
\ee
Note that $\hat{u}_\theta(x_{TS})$ is  the unit vector tangent to the dominant path at the configuration $x_{TS}$ ~($l_{TS}$ is the value of the curvilinear abscissa a for which Eq. (\ref{transition}) is satisfied).
Correspondingly, the partition function reads 
\be
\label{ZTS}
Z_{\partial W} \equiv Z_{TS} = {\int d x ~ \delta\left[ (x-x_{TS})\cdot \hat u_{\theta}(x_{TS}) \right]~e^{-\beta U(x)} },
\ee

The partition function (\ref{ZTS}) can be estimated in local harmonic approximation, by running short MD 
simulations starting from $x_{TS}$, subject to the constraint to lie on the surface 
orthogonal to the tangent to the reaction pathway at $x_{TS}$, i.e. 
\be
Z_{TS}  \simeq ~e^{-\beta U(x_{TS})}~\prod_{i=1}^d~ \sqrt{2 \pi \langle (x^i-x^i_{TS})^2\rangle_\perp},
\ee
where $x^i$ is the $i-$th coordinate of the configuration $x$ and
\be
\langle (x^i-x^i_{TS})^2 \rangle_\perp \equiv    \frac{\int d x ~(x^i-x^i_{TS})^2~ \delta\left[ (x-x_{TS})\cdot \hat u_{\theta} \right]
~e^{-\beta U(x)}}{\int d x~ \delta\left[ (x-x_{TS})\cdot \hat u_{\theta} \right]
~e^{-\beta U(x)}}.
\ee

Hence, using the DRP expression for the reaction current we arrive to our final result:
\be
\label{final1}
k_{R\to P} &\simeq& Z_{TS}~ \left| \sqrt{\frac{E_{eff}+V_{eff}(x_{TS})}{D}} \hat u_\theta(x) -\frac{\beta}{2} \nabla U(x_{TS}) \right|
  ~\frac{P_{reg.}(x_0, t|x_0)]}{e^{E^{reg.}_{eff} t -  S_{HJ}^{reg.}[\ox_{reg.}]}} ~\sqrt{\det\left( \hat{F}_{reg.}[\ox_{reg.}]\hat{F}^{-1}[\ox]\right)}~e^{-\frac{\beta}{2 }(U(x_{TS})-U(x_i)) + E_{eff} t[\ox] - S_{HJ}[\ox] }.\nonumber\\
\ee

In such an Eq., the effective energy parameter $E_{eff}$ must be chosen in such a way that the total time $t[\ox]$ evaluated according to 
Eq. (\ref{time}) is much larger than the relaxation time in the reactant and yet much smaller than the total relaxation rate. In such a time regime, the flux of probability current across the transition state is stationary and the value of the expression (\ref{final1}) must be independent on the specific value of $E_{eff}$ chosen. 
 
Finally, if the reaction can occur through more than one reaction pathway,  the total rate is obtained simply by adding up all such contributions:
\be
\label{final2}
k_{R\to P} &\simeq& \sum_k Z_{TS}^k~e^{\beta U(x^k_{TS})}~ |J(x_{TS}^k, t|x_i)|.
\ee

For very simple systems, computing the rate by means of  Eq.(\ref{final1}) of Eq. (\ref{final2}) is expected to be more computationally  expensive than in standard Kramers theory~\cite{review}. 
On the other hand, the advantage of the DRP method developed here is that it does not require to know \emph{a priori} the location of the transition state. Hence, we expect that our method 
may  be used to investigate the kinetics of two-state reactions in large configuration spaces, which are generally characterized by a complicated energy surface. 
In particular, for many complex molecular systems, the transition state cannot by guessed from the structure of the interaction,  and the multi-dimensional Kramers theory is therefore useless. 

We also observe that the DRP method presented in this section bares some similarity with other existing techniques for rate calculation. In particular, the identification of the reaction coordinate $l$ from 
a statistical important reaction pathway is also used in the so-called milestoning method~\cite{milestoning}. 
An important advantage of such an approach with respect to the DRP method developed here is that it does not require to assume two-state kinetics. On the other hand, the present method is much less computationally expensive, as it does not require to evaluate the first-passage-time distributions from the different milestones, from MD simulations. 

The DRP Eq. (\ref{final1}) bares also some similarity also with Chandler's theory~\cite{chandler} and with transition state theory~\cite{tst}. Indeed, in  both such approaches, the rate is related to  the flux of reactive
 trajectories through the transition state. On the other hand, we stress the fact that in the present DRP approach, such a  flux is evaluated over non-equilibrium trajectories. 

 Finally we note that, in the transition path sampling algorithm~\cite{TPS2}, the rate is usually calculated starting from the path integral expression of the product population fraction, i.e. Eq. (\ref{nPPI}). 
 However, in such an approach, the normalization of the path integral is obtained by evaluating the free energy of the transition path ensemble, i.e. the reversible work  which is required to constraint the final
  configurations of the  paths into the product state. On the other hand, in the DRP approach, such a normalization  is guaranteed by construction  \emph{a priori},  by the regularization procedure. 
  
\section{Testing the DRP Calculation of the Reaction Rates}
\label{testrate}
In order to test the scheme developed in the previous section for rate calculations, we study the reaction kinetics of some simple two-dimensional toy systems, for which very accurate results can
be obtained also using other methods. In particular, in wide range of temperatures, the rate for such systems 
can be accurately evaluated using Kramers theory or  computed directly by running long MD simulations.

\begin{figure}[t]
\includegraphics[width=8.5 cm]{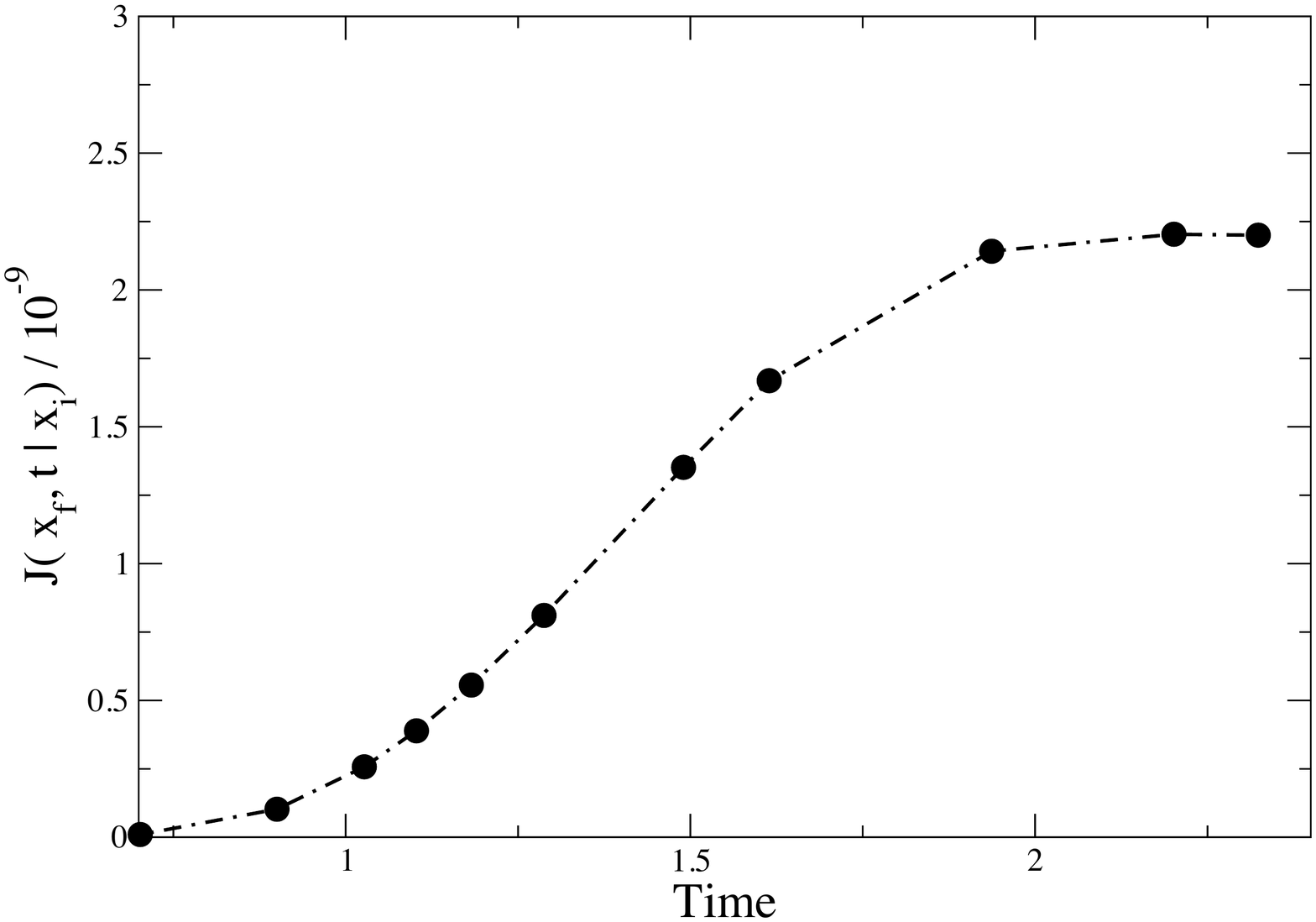}
\includegraphics[width=8.5 cm]{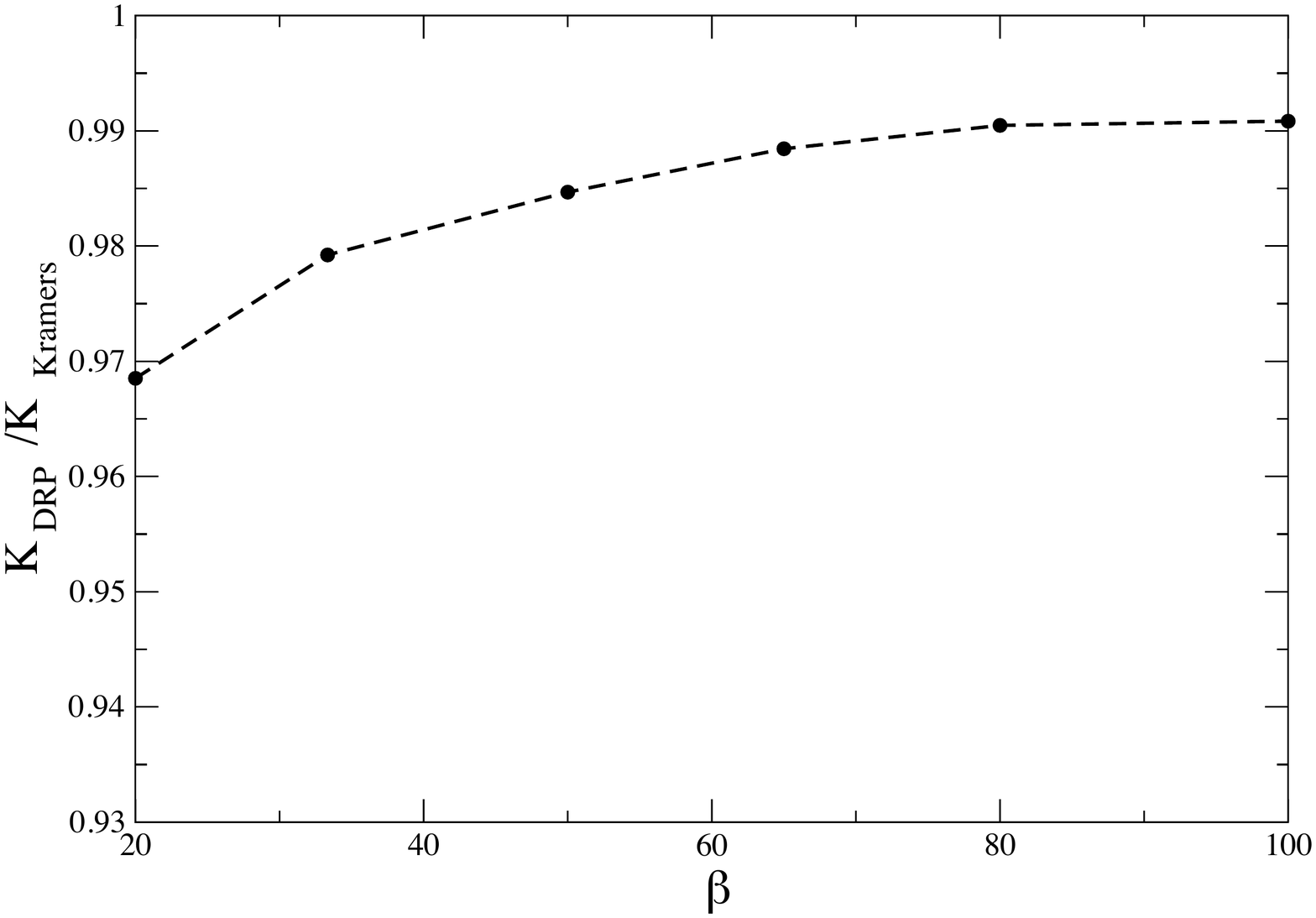}
\caption{Left panel: The absolute value of the normalized probability current at the sadde-point $x_f=0$, evaluated with the DRP method at different times $t$, for $x_i=-\omega$ and $\beta=5$. The different total times are determined by different choices of the effective energy parameter. 
Righ panel: The ratio between the reaction rates evaluated with the DRP method and with Kramers theory.}
\label{current}
\end{figure}

Let us begin by considering the rate of escape from a standard two-dimensional bi-stable potential,
\be
U(x,y) =  a  (x^2 - \omega^2)^2 + b y^2
\ee 
with $a=1/8$, $b=2$,  and $\omega = 2$. 
In this simple case, the dominant path connecting $x_i=-\omega$ to $x_f= \omega$ is known analytically (straight horizontal line). We have represented such a path using $N_s=150$ equally spaced
steps.
In addition, performing an accurate calculation of the fluctuation determinants for such a simple two-dimensional system is straightforward, even without residing on the low time resolution effective
description.  

We recall that in the DRP approach, the total time of the transition is determined by the value of the effective energy parameter $E_{eff}$,  according to Eq. (\ref{time}).
The same parameter is used in the calculation of all the discretized time intervals which enter in the fluctuation determinant.
On the other hand, our prediction for the rate must obviously not depend on the choice of $E_{eff}$.

In order to see how this condition can be satisfied, we recall that our rate expression (\ref{shorttK}) requires the time interval $t$ to be much smaller than the inverse rate, $k_{R \to P} t \ll 1$. 
At the same time, $t$ must be chosen  much larger than 
the thermalization time in the reactant state, to assure single exponential relaxation.
If both conditions are simultaneously satisfied, then one should observe a stationary flux through the transition state. Consequently, our rate expression (\ref{final2}) 
should become independent on the precise choice of $t$ --- hence of $E_{eff}$---.

In order to test if such a stationary current is realized in our simulations, in the left panel of Fig. \ref{current} we plot the DRP current at the saddle-point as a function 
of the total time (hence for different values of the effective energy). We can clearly see that for the longest times the current becomes stationary, 
hence the rate stops depending on the specific choice of $E_{eff}$. 
The right panel of Fig.~\ref{current} shows that, in this system,  the calculation of the rate using Kramers' theory using the DRP approach  agree 
within $\simeq 1 \%$ accuracy.

\begin{figure}[t]
\includegraphics[width=6.5 cm]{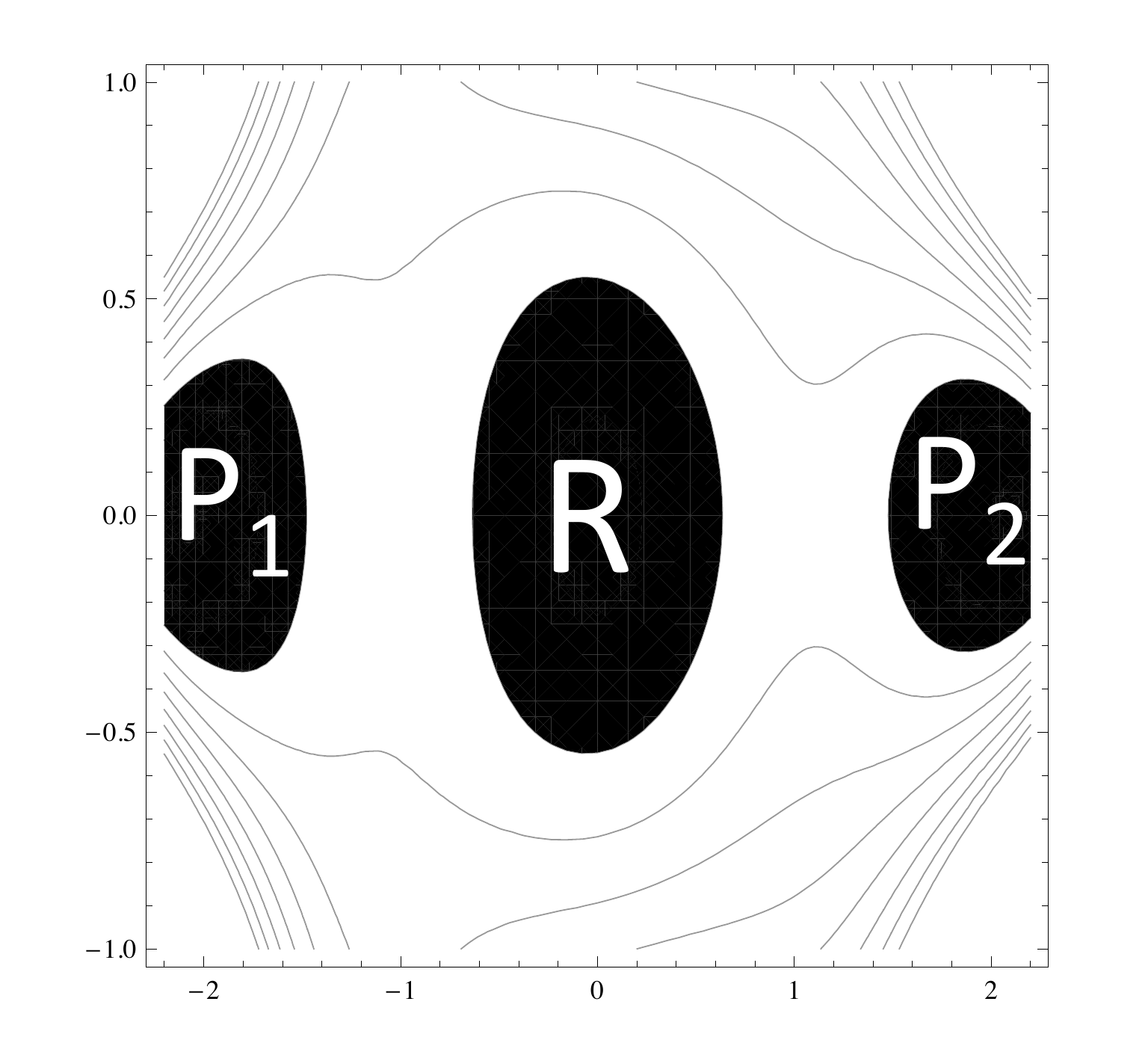}\qquad
\includegraphics[width=8.5 cm]{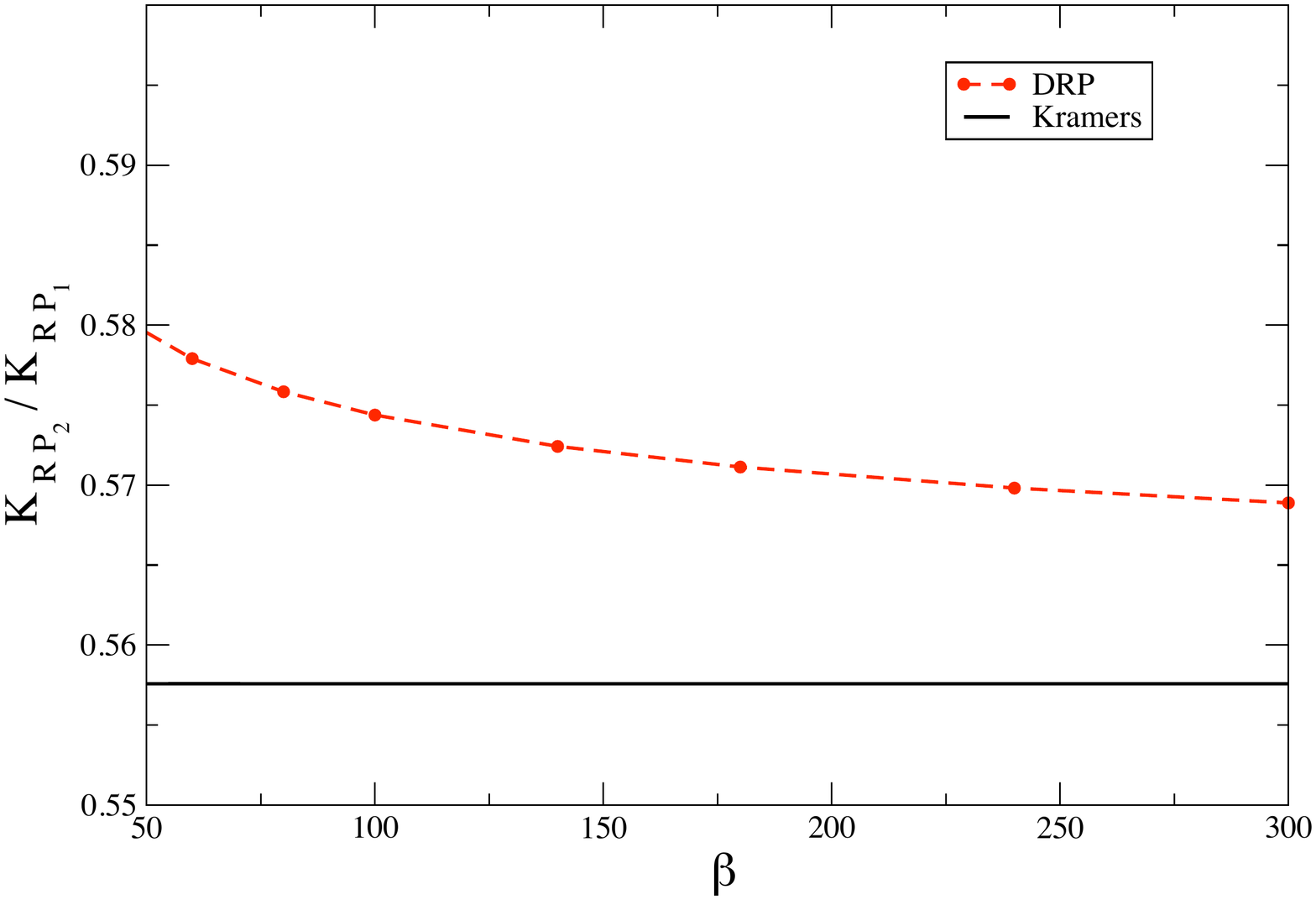}
\caption{Left panel: The contour plot of the potential energy (\ref{Upot2}). Right panel: the ratio of the rates $k_{R\to P_2}/k_{R\to P_1}$ evaluated in the DRP approach and using Kramers theory.  
using Kramers theory, DRP calculations and direct MD simulations. }
\label{3Dcase}
\end{figure}

Let us now consider a  toy model which allows to specifically  assess  the role played by the stochastic fluctuations around the dominant path, in the kinetics of  the reaction.
To this end, we study the two-dimensional three-state system, consisting of a reactant $R$, and of two products $P_1$ and $P_2$, shown in the left panel of Fig. \ref{3Dcase}
The functional form of the potential energy is
\be
\label{Upot2}
U[x, y] :=  a x^6 + b x^4 + c  x^2 +   k_1 \exp\left[-\frac{(x - x_1)^2}{2 \sigma}\right]  + (x^4 -  2 x^2 + \kappa x + \Omega) y^2
\ee
with $a=5/64$, $b=-10/16$, $c=1$, $k_1=-0.6$,  $x_1=0$, $\sigma=0.3$, $\kappa=0.5$ and $\Omega=2$. Also in this case, the dominant paths are known analytically. The HJ action 
and the fluctuation determinants appearing in Eq. (\ref{final1}) are evaluated using $N_s= 100$ discretization steps.

This model is built is such a way that the energy barrier separating the reactant to the two products is the same. Yet, the steeper structure of the potential energy 
on the right of the reactant tends to disfavor large stochastic fluctuations, along the dominant path of the reaction $R\to P_2$. We therefore expect that the rate of escape from $R$ to $P_1$ should be 
larger
than that from $R$ to $P_2$. 
This fact is clearly seen in the right panel of Fig. \ref{3Dcase}, were we compare the ratio $k_{R\to P_1}/k_{R\to P_2}$ evaluated using Kramers theory and 
using the DRP method. We see that the two approaches give results which agree within about  $2\%$ accuracy. In both cases, the reaction $P\to P_2$ is about $50\%$ slower than the reaction $P\to P_1$.

\section{Conclusions}
\label{conclusions}
In this paper, we have analyzed the role of the stochastic fluctuations around the most probable reaction pathways, in systems obeying the over-damped Langevin dynamics. 
Such fluctuations affect the probability for a transition to take place through a specific reaction channel in a given time, hence the kinetic of the reaction. 
For sufficiently small temperatures,  the fluctuations in the ensemble of reaction pathways are confined within small bundles around the locally most probable 
reaction pathways.  In such a regime, we have developed a technique to efficiently compute their contribution, to order $k_B T$ accuracy. 
Clearly, in the opposite high-temperature regime, the stochastic fluctuations become very large and overshadow the information encoded in the dominant paths. In this case, the accuracy of the
DRP approach breaks down. 
 
 We have shown that the calculation of the time intervals and of the fluctuation determinant
  can be made much more efficient by adopting a low-time resolution effective description, in which the fast dynamics is analytically pre-averaged out.
 Indeed, in the effective stochastic theory, the convergence of DRP calculations is achieved using much fewer discretization paths.   
 
  The $o(k_B T)$ expression of the conditional probability was used to derive a formula  for the reaction rate, within a flux-over-population approach. 
We have illustrated and tested our results on simple toy models, where accurate or even exact results could be obtained with alternative methods. 
 In the future, we plan to apply these techniques to investigate the kinetics of much more complicated molecular reactions. 
 
\acknowledgments
Part of this work was performed when P.F. was visiting the IPhT at CEA (Saclay) under a CNRS grant. 
P.F. and S.B. are members of the Interdisciplinary Laboratory for Computational Sciences (LISC), a joint venture of Trento University and FBK.  The authors acknowledge useful discussions with F. Pederiva.
\appendix

\section{The Conditional Probability for the Diffusion in a $d-$Dimensional  Harmonic Potential}
\label{PHOapp}
Here we review the exact calculation of the conditional probability for a point-particle diffusing in the $d-$dimensional harmonic external potential 
\be
U(x) = \frac{1}{2} \sum_i \omega_i^2 (x^i-x^i_0)^2.
\ee
For sake of simplicity, we discuss such an evaluation  in the case $x_f=x_i=0$ and we choose the origin of the configuration space in such a way $x_0^i=0$. We also choose units in which the viscosity
$\gamma = 1/\beta D$ is set to 1.  

In this case the effective action of the static solution $\ox=0$ is $S_{eff}[\ox] = -\frac{t}{2} \sum_i ~\omega_i^2$, hence:
\be
P_{HO}(0,t|0) &=& \mathcal{N} e^{ \frac{t}{2}~\sum_i ~\omega_i^2 t} 
\frac{1}{\sqrt{\det\left[\delta_{i j} (-\frac{\beta}{2}\frac{\partial^2}{\partial t^2} + \frac \beta 2 \omega_i^4)\right]}}\nonumber\\
&=&\mathcal{N'} e^{ \frac{t}{2}~\sum_i ~\omega_i^2 t}~\left(\frac{2}{\beta}\right)^{d/2} 
\frac{1}{\sqrt{\det\left[\delta_{i j} (-\frac{\partial^2}{\partial t^2} +  \omega_i^4)\right]}},
\ee
where in the last step we have absorbed the factor $\beta/2$ which appears in the fluctuation operators into the normalization constant $\mathcal{N}$.

The normalized eigen-functions and eigen-values of the fluctuation operator with the boundary conditions $y(t)=y(0)=0$ are:
\be
y^i_n(\tau) &=& \sqrt{\frac{2}{t}} \sin \left[ \frac{n~\pi}{t} \tau\right]\\
\lambda^i_n &=&  \frac{(\pi n)^2}{ t^2} + \omega_i^4 \qquad (n=1,2,\ldots, \quad i=1,\ldots,d)
\ee
Notice that the same eigenfunctions are also eigenstates of the fluctuation operator for the free diffusion
\be
\hat F_{0} = - \delta_{i j} \delta(\tau' - \tau)~\frac{ \partial^2}{\partial \tau^2}, 
\ee
with eigenvalues
\be
\lambda^i_{0\, n} &=&  \left(\frac{\pi }{t}\right)^2 n^2 \qquad (n=1,2,\ldots, \quad i=1,\ldots,d)
\ee

As usual, in order to get rid of the unknown normalization $\mathcal{N}'$ we multiply and divide by a regulator, in this case the free propagator
$P_0(0,t|0)$:
\be
P_{HO}(0,t|0) &=& P_0(0,t|0) ~e^{ \frac{t}{2}~\sum_i ~\omega_i^2 } 
~\sqrt{\frac{\prod_{i=1}^{d}~\prod_{n\ge1 }\left(\frac{(n \pi)^2}{ t^2 }\right)} 
{\prod_{i=1}^{d}~\prod_{n\ge 1}^\infty \left(\frac{(n \pi)^2}{ t^2 } +\omega_i^4\right) }}\\
&=& \left(\frac{\beta}{4 \pi ~t}\right)^{d/2}~e^{ \frac{t}{2}~\sum_i ~\omega_i^2 }  
\sqrt{\frac{1} {\prod_{i=1}^{d}~\prod_{n\ge 1}^\infty \left(1 + \frac{\omega_i^4 t^2}{(n^i \pi)^2}\right)}}
\ee

Using the result
\be
\prod_{n\ge 1} \left(1 + \frac{\omega_i^4 t}{(n^i \pi)^2}\right) = 
\frac{\textrm{Sinh}( \omega_i  t )}{ \omega_i^2~t}
\ee
we find
\be
P_{HO}(0,t|0) &=& \left(\frac{\beta}{4~\pi }\right)^{d/2}~e^{ \frac{t}{2}~\sum_i ~\omega_i^2 } 
\prod_{i=1}^{d}\sqrt{\frac{\omega_i^2}{~ \textrm{sinh}(\omega_i^2 t)}}
\ee

Notice that, in the long time limit, $ \textrm{sinh}(\omega_i^2 t)\to \frac{1}{2}e^{\frac{\alpha_i t}{2}}$, so $P_{HO}(0,t|0)$ converges to the inverse of the partition function, as it should:
\be
P_{HO}(0,t|0) \to \left(\frac{\beta}{2\pi }\right)^{d/2}~\prod_i \omega_i = \frac{1}{Z_{HO}}
\ee
Notice also that, for this system, the thermalization time does not depend on the temperature. 

\section{Effective Stochastic Theory}
\label{ESTapp}

In this section, we sketch the derivation of the  EST. For all further details we refer the reader to the original paper \cite{EST}. 
 For simplicity and without loss of generality, it is convenient to consider the path integral with periodic boundary conditions    
\be
Z(t) \equiv \int d x ~P(x| x;t) =  \oint \mathcal{D} x ~e^{- S_{eff}[x]}.
\label{pPI}
\ee
The starting point to develop the EST consists in introducing the Fourier components of the paths, 
\be
\tilde x(\omega_n) &=& \frac{1}{t}~\int_0^t d\tau ~x(\tau)~e^{- i \omega_n t }  \\
x(\tau) &=& x(\tau+t) =  \sum_n x(\om_n) ~ e^{ i \omega_n t }.
\ee
where $\om_n \frac{2 \pi}{t}~n$,  are the Fourier  frequencies $( n=0, \pm 1, \pm 2, \ldots)$.

The path integral (\ref{pPI}) is defined in the continuum limit. Numerical simulations are always performed using a finite discretization time step $\Delta t$. 
Clearly, the shortest time intervals which can be explored in a numerical simulation is of the order of few  $\Delta t$. Equivalently,  the largest frequencies of the  Fourier transform of the stochastic paths $\tilde x(\omega)$ are of the order few fractions of an ultra-violet (UV) cut-off $\Omega \equiv 2 \pi/ \Delta t.$
 
Let us now split the Fourier modes of the paths contributing to (\ref{PI})  in high-frequency ---or \emph{"fast"}--- modes and low-frequency ---or \emph{"slow"}--- modes. To this end, 
we introduce  a real number  $0<b<1$   such that the 
frequency  range $(0, \Om)$ is split in two intervals $(0, b~\Omega)\cup~(b~\Om, \Om)$. Correspondingly, one can define the "fast" component of the path $x_>(\tau)$ and the "slow" component of the path $x_<(\tau)$, by summing over the Fourier modes in the  $(0, b~\Omega)$ and $(b~\Om, \Om)$ range, respectively: 
 \be
x_<(t) &=& \sum_{|\om_n| \le b\Om}~\tilde x(\omega_n) ~e^{~i \om_n t }\\
 x_>(t) &=& \sum_{b\Om\le |\om_n| \le \Om}~\tilde x(\omega_n) ~e^{~i \om_n t }.
  \ee

The complete path integral  (\ref{pPI}) can therefore be exactly re-written in the following way:
\be
\label{effAct}
Z(t)~=  \oint   \mathcal{D}x_<\oint \mathcal{D}x_>~ e^{-S_{eff}[x_<+x_>]}&\equiv& \oint  \mathcal{D}x_<~ e^{-S_{eff}[x_<]}~e^{- S_>[x_<]},
\ee
where 
\be
\label{path2}
e^{- S_>[x_<(\tau)]}\equiv \oint \mathcal{D}x_> e^{S_{eff}(x_<] - S_{eff}[x_<+x_>]}
\ee
is called the renormalized part of the effective action. 

The EST is constructed by explicitly evaluating $S_{>}[x_<]$, i.e.  by performing the path integral over fast modes $x_>(\tau)$. In the limit in which the fast and slow modes are separated by a large gap in the spectrum of Fourier modes --- i.e. if the system displays a decoupling of time scales---such an integral can be carried out analytically in a perturbative approach based on Feynman diagram techniques\cite{EST}. 
The expansion parameter such a perturbation theory is the ratio between the typical frequency $\omega$ of the slow modes  and the UV cut-off  $b\Omega$.
Clearly, if hard and slow modes are decoupled, the ratio $\omega/(b\Omega)$ is a small number, hence the terms proportional to higher and higher powers  $L$ of  such a ratio  provide smaller and smaller corrections. 

If one accounts only for the leading corrections in the $1/b\Omega$  expansion, the renormalized part of the action takes the form of an effective interaction term~\cite{EST}, i.e. 
\be
e^{- S_>[x_<(\tau)]} = e^{-\int_0^t d\tau ~V_{eff}^R[x_<(\tau)]}
\ee
where 
\be
V^{R}_{eff} (x) &\simeq& 
\frac{D_0~(1-b)}{~\pi~ b\Omega} ~\nabla^2 V_{eff}(x).
\ee   

We emphasize that the result of the EST construction is a new expression for the \emph{same} path integral (\ref{pPI}), in which the UV cutoff been lowered from $\Omega$ to $b \Omega$. Equivalently, the  path integral is discretized according to a  larger elementary  time step,  $ \Delta t \to \Delta t/b$: 
\be
Z^{\Delta t}(t) &\equiv& \oint_{\Delta t} \mathcal{D} x ~e^{- S_{eff}[x]} \propto \oint_{\Delta t/b} \mathcal{D} x ~e^{- S_{eff}[x] - 
 \int_0^t d\tau ~V^R_{eff}[x(\tau)]} \equiv Z^{\Delta t/b}_{EST}(t)
\label{EST}
\ee
In these expressions, the symbol $\oint_{\Delta t}$ denotes the fact that the path integral is discretized according to an elementary time step $\Delta t$ and we have suppressed the subscript "<",  in the paths. It can be shown that the proportionality factor between $Z^{\Delta t}(t)$ and $Z^{\Delta t/b}_{EST}(t)$ depends only on $t$ and does not
contribute to the statistical averages.

\section{Path Integral Expression for the Population Fractions $n_P(t)$ and $n_R(t)$}
\label{PInP}

In this section we provide a microscopic representation of the solutions $n_P(t)$ and $n_R(t)$ of the kinetic Eq.s (\ref{rkin1})-(\ref{rkin2}). 
Let us consider in particular the case in which the system is initially prepared in the reactant state, i.e. $n_R(0) =1$ 
and $n_P(0)= 0.$ Let $\rho_0(x)$ be the initial distribution of the configurations in the reactant state. Clearly, such a choice of initial conditions implies the normalization condition
\be
\int d x \ h_R(x) \rho_0(x) = 1,
\ee
where $h_R(x)$ is the characteristic function of the reactant, i.e. $h_R(x)=1$ if $x\in R$ and $0$ otherwise. 
Under the assumption of two state kinetics,  the thermalization in the reactant occurs over a very short time scale, hence the specific choice of the initial distribution in the reactant is in fact irrelevant. 

We now show that a microscopic representation for the product population fraction $n_P(t)$  is obtained by \emph{averaging} the conditional probability $P(x_f,t|x_i)$ over the initial configurations $x_i$ in the reactant and \emph{summing} over all possible final configurations $x_f$ in the product, i.e.  
\be
n_{P}(t) &=& \int d x_f \ h_P(x_f) \int d x_i \ h_P(x_i) \ \rho_0(x_i)~P(x_f,t|x_i)\nonumber\\
&=& ~\mathcal{N}~\int d x_f \ h_P(x_f) \int d x_i \ h_P(x_i) \rho_0(x_i)~ e^{-\frac \beta 2 (U(x_f)-U(x_i))} ~\int_{x(t_i)=x_i}^{x(t)=x_f} \mathcal{D} x~e^{-\int_{0}^t d\tau 
\left(\frac{\dot{x}^2}{4 D} + V_{eff}[x]\right)}.
\label{RtoP}
\ee 
We observe that Eq. (\ref{RtoP})  satisfies the correct initial condition,  $n_P(0)=0$. 

Let us now introduce  the complete set of eigenstates $\Psi_n(x)$ of the "quantum" Hamiltonian $\hat H_{eff}$, 
 \be
 \hat H_{eff} \Psi_n(x) = k_n \Psi_n(x). 
 \ee 
 In particular, it is immediate to verify that the ground state of $\hat H_{eff}$ has a vanishing eigenvalue and reads
 \be
 \Psi_0(x)  = \frac{e^{-\frac{\beta}{2} U(x)}}{\sqrt{Z}}, 
 \ee
 where $Z$ is the partition function of the system,
 \be
 Z = \int d x\  e^{-\beta U(x)}.
 \ee
 
By inserting the resolution of the identity, $1 = \sum_n | n\rangle ~\langle n |$, into  the "quantum" propagator (\ref{K}) we obtain the so-called 
spectral representation of the conditional probability:
\be
\label{spec}
P(x,t|x_i) = e^{-\beta/2 (U(x)-U(x_i))} ~\sum_{n=0}^{\infty}~ \Psi^*_n(x)~\Psi_n(x_i) e^{-k_n t}.
\ee
Hence, the conditional probability $P(x_f, t|x_i)$ converges to the Boltzmann distribution, in the long time limit:
\be
\label{equi}
P(x,t|x_i)~\stackrel{t\to\infty} {\rightarrow}~\frac{1}{Z}e^{-\beta U(x)},
\ee
regardless of the initial condition. 

In particular, the systems obeying two-state kinetics are those in which the spectrum displays a gap between the first and second eigenstates of the effective "quantum" Hamiltonian $H_{eff}$:
\be
\label{2s}
k_1\ll k_2.
\ee
Indeed, in this case  the time scale $\tau_1 =\frac {1}{ k_1}$ decouples from all the other relaxation time scales in the system, and the approach to thermal equilibrium occurs through a single-exponential relaxation:
\be
\label{relaxation}
P(x,t|x_i) \simeq \frac{1}{Z}e^{-\beta U(x)} +  e^{-\beta/2 (U(x)-U(x_i))}~ \Psi^\dagger_1(x) \Psi_1(x_i) ~ e^{- k_1 t}.  
\ee
If the reaction is two-state and if $x_i$ and $x_f$ are not in the same state (for example $x_i\in R$ and $x_f\in P$) then the probability of performing a transition from $x_i$ to $x_f$ vanishes in the short-time limit, i.e. 
\be
\label{tto0}
\lim_{t\to 0}~ P(x_f, t| x_i)= 0.
\ee
This fact implies that
\be
\Psi^\dagger_0(x_f) \Psi_0(x_i) = -\Psi^\dagger_1(x_f) \Psi_1(x_i),
\label{rule}
\ee
thus Eq. (\ref{relaxation}) gives
\be
P(x_f,t|x_i) = \frac{e^{-\beta U(x_f)}}{Z}~\left( 1 - e^{-k_1 t} \right). 
 \ee
We emphasize that Eq. (\ref{tto0}) ---and therefore Eq. (\ref{rule})--- are \emph{not} justified if $x_f$ and $x_i$ are in the same state. Indeed, in the approximation of two-state kinetics, the local thermalization in the $R$ and $P$ states is assumed to occur instantaneously, as the only finite time scale is the mean-first-passage time across the barrier.

Using  the spectral decomposition (\ref{relaxation})  we find
\be
n_{P}(t) &=& \int d x_f \ h_P(x_f) \int d x_i \ h_R(x_i) \ \rho_0(x_i)~ \frac{ e^{-\beta U(x_f)}}{Z}~\left( 1-  e^{- k_1 t} \right).
\ee
The product population fraction at time $t$ then reads
\be
n_{P}(t) = n_P^{eq} \left( 1-  e^{- k_1 t} \right).
\ee
where $n_P^{eq}= \frac{Z_P}{Z}$. Hence, we have recovered Eq. (\ref{solrkin1}) and we have shown that first excited state of the quantum effective Hamiltonian is the equilibrium relaxation rate  of the system, $k_1= k$.


\begin{thebibliography}{99}
\bibitem{TPS1}  C.~Dellago.  P.~G. Bolhuis, F.~S. Csajka, and  D.~Chandler, J. Chem. Phys. {\bf 108}, 1964 (1998).  
\bibitem{TPS2} P.~G. Bolhuis,  D.~Chandler,  C.~Dellago, and  P.~L. Geissler, Ann. Rev. Phys. Chem. {\bf 53}, 291 (2002).
\bibitem{Elber} A.~Ghosh, R. Elber and H.~A. Sheraga, Proc. Nat. Acad. Sci. {\bf 99}, 10394 (2002). 
\bibitem{DIMS} D.M.~Zuckerman and T.B. Woolf, Phys. Rev. {\bf E 63}, 016702 (2000).
\bibitem{elber1} R.~Elber,  and D.~Shalloway, J. Chem. Phys. {\bf 112} 5539 (2000).
 \bibitem{DRP1} P.~Faccioli, M.~Sega, F.~Pederiva and H.~Orland,  Phys. Rev. Lett. {\bf 97}, 108101  (2006). 
\bibitem{DRP2}  M.~Sega, P.~Faccioli,  F.~Pederiva, G.~Garberoglio and H.~Orland, Phys. Rev. Lett. {\bf 99}, 118102 (2007). 
\bibitem{DRP3}  E.~Autieri, P.~Faccioli, M.~Sega, F.~Pederiva and H.~Orland,  J. Chem Phys. {\bf 130}, 064106 (2009).
\bibitem{Donniach} P.~Eastman, N.~Gronbech-Jensen, and S.~Doniach, J. Chem. Phys. {\bf 114}, 3823 (2001).
\bibitem{DRPappl1} S.~a Beccara, P.~Faccioli, G.~Garberoglio, M.~Sega, F.~Pederiva, H.~Orland,   arXiv:1007.5235, J. Chem. Phys. in press
\bibitem{DRPappl2}   S.~a Beccara, G.~Garberoglio,  P.~Faccioli and F.~Pederiva,  J. Chem. Phys. {\bf 132} 111102 (2010) (comm.)
\bibitem{DRPtest1}   P.~Faccioli,   Journ. Phys. Chem. {\bf B112} (2008) 13756. 
\bibitem{DRPtest2} P.~Faccioli, A. Lonardi and H. Orland,  J. Chem. Phys. {\bf 133}, 045104 (2010).
\bibitem{EST} O.~Corradini, P.~Faccioli and H.~ Orland, Phys. Rev. {\bf E80} (2009) 061112.
\bibitem{RGMD} P. Faccioli,  J. Chem. Phys. {\bf 133} 164106(2010)
\bibitem{review} P. H\"anggi, P. Talkner and M. Borkovec, Rev. Mod. Phys. {\bf 62}, 251 (1990).   
\bibitem{milestoning} A. K. Farajian and R. Elber, J. Chem. Phys. {\bf 120}, 10880 (2004). 
\bibitem{chandler} D.~Chandler, J. Chem. Phys. {\bf 68}, 2959 (1978).
\bibitem{tst} D.G.~Truhlar, B.C.~Garrett and S. J.~Klippenstein, Journ. Phys. Chem. {\bf 100}, 12771 (1996).


\end{thebibliography}
 \end{document}